\documentclass[10pt,final,journal,letterpaper,oneside,twocolumn]{IEEEtran}
\usepackage{physics}
\usepackage{amsmath}
\usepackage{amsfonts}
\usepackage{mathtools}
\usepackage{amsfonts}
\usepackage{makecell}
\usepackage{amssymb}
\usepackage{hyperref}
\usepackage{array}
\usepackage{makeidx}
\usepackage{longtable}
\usepackage{csquotes}
\usepackage{graphicx}
\usepackage{cite}
\usepackage{xcolor}
\usepackage{subcaption}
\usepackage{graphicx}
\usepackage{multicol}
\usepackage{multirow}
\usepackage{color,soul}
\usepackage{pifont}
\usepackage{booktabs}
\usepackage[acronym]{glossaries}

\begin{document}
%

\title{\huge RIS-based Physical Layer Security for Integrated Sensing and Communication: A Comprehensive Survey}
%
%
%

\author{  Yongxiao~Li, Feroz Khan, Manzoor~Ahmed,
        Aized Amin Soofi,
         Wali Ullah Khan,~\IEEEmembership{Member,~IEEE,} Chandan~Kumar~Sheemar~\IEEEmembership{Member,~IEEE,}, Muhammad Asif      
        and Zhu~Han,~\IEEEmembership{Fellow,~IEEE}
\thanks{Yongxiao Li and Feroz Khan are with the School of Electronic Engineering, Beijing University of Posts and Telecommunications, Beijing, China (e-mails:  ferozkhan687@gmail.com, Yongxiao.li@bupt.edu.cn).} 
\thanks{Manzoor Ahmed, is with the School of Computer and Information Science and also with Institute for AI Industrial Technology Research, Hubei Engineering University, Xiaogan City 432000, China  (e-mails: manzoor.achakzai@gmail.com),
}
\thanks{Aized Amin Soofi is with the Department of Computer Science, National University of Modern Languages Faisalabad, 38000, Pakistan (e-mail:aizedamin@gmail.com ).}

\thanks{Wali Ullah Khan and Chandan Kumar Sheemar are with the Interdisciplinary Centre for Security, Reliability, and Trust (SnT), University of Luxembourg, 1855 Luxembourg City, Luxembourg (e-mails: waliullah.khan@uni.lu, chandankumar.sheemar@uni.lu).}

\thanks{Muhammad Asif is with the School of Computer Science and Communication Engineering, Jiangsu University, Zhenjiang, China (e-mail: masif@ujs.eu.cn)}
\thanks{Zhu Han  is with the Department of Electrical and Computer Engineering at the University of Houston, Houston, TX 77004 USA, and also with the Department of Computer Science and Engineering, Kyung Hee University, Seoul, South Korea, 446-701 (e-mail: hanzhu22@gmail.com).}
}

%
%

\markboth{Submitted to IEEE IoT}{Feroz Khan \MakeLowercase{\textit{et al.}}: RIS-PLS-ISAC}

%



\maketitle
\begin{abstract}
Integrated Sensing and Communication (ISAC) is a crucial component of future wireless networks, enabling seamless integration of Communication and Sensing (C\&S) functionalities. However, ensuring security in ISAC systems remains a significant challenge, as both C\&S data are susceptible to adversarial threats. Physical Layer Security (PLS) has emerged as a key framework for mitigating these risks at the transmission level. Reconfigurable Intelligent Surfaces (RIS) further enhance PLS by dynamically shaping the radio environment to improve both secrecy along with C\&S performance.
This survey begins with an overview of RIS, PLS, and ISAC fundamentals, establishing a foundation for understanding their integration. The state-of-the-art RIS-assisted PLS approaches in ISAC systems are then categorized into passive RIS and Active RIS (ARIS) paradigms. Passive RIS-based techniques focus on optimizing system throughput, covert communication, and Secrecy Rates (SRs), alongside improving sensing Signal-to-Noise Ratio (SNR) and Weighted Sum Rate (WSR) under various constraints. ARIS-based strategies extend these capabilities by actively optimizing beamforming to enhance secrecy and covert rates while ensuring robust sensing under communication and security constraints.
By reviewing both passive and ARIS-based security frameworks, this survey highlights the transformative role of RIS in strengthening ISAC security. Furthermore, it explores key optimization methodologies, technical challenges, and future research directions for integrating RIS with PLS to ensure secure and efficient ISAC in next-generation 6G wireless networks.

\end{abstract}


%
\IEEEpeerreviewmaketitle

%
%
%
%
\section{Introduction}

Integrated Sensing and Communication (ISAC) has emerged as a cornerstone technology for future wireless networks, particularly in the evolution of Sixth Generation (6G) communication systems \cite{mandelli2023survey, lu2024integrated, singh2025isac}. By seamlessly merging Communication and Sensing (C\&S) functionalities within a unified framework, ISAC not only facilitates efficient spectrum utilization but also optimizes resource management. This integrated approach reduces hardware redundancy, significantly enhancing overall system efficiency \cite{volgushev2024integrated}. Moreover, the fusion of these capabilities is essential for supporting a wide range of next-generation applications, such as autonomous driving, smart healthcare, industrial automation, digital twins, and Vehicle-to-Everything (V2X) communication. ISAC architectures can be customized to be communication-centric, sensing-centric, or co-designed, with each approach meticulously balancing communication throughput and sensing accuracy based on specific system requirements\cite{zhu2023pushing, dai2022survey}.

A fundamental advantage of ISAC lies in its ability to leverage existing communication infrastructure for large-scale sensing. This capability eliminates the need for dedicated radar networks, allowing wireless signals to simultaneously transmit data while extracting valuable environmental information. However, the integration of sensing and communication introduces inherent security risks, as both data transmission and sensing outputs become vulnerable to eavesdropping, jamming, and spoofing attacks. Addressing these concerns is essential to ensure the confidentiality, integrity, and reliability of ISAC-enabled systems \cite{singh2025integrated}. The dual functionality of ISAC systems poses significant security challenges. Adversaries may exploit transmitted signals to intercept confidential data or manipulate sensing outcomes.
Unlike conventional wireless networks, ISAC interlinks the mutual information between sensing and communication, making it more susceptible to unauthorized access, data leakage, and adversarial interference. These threats can severely disrupt critical applications such as autonomous navigation, defense systems, and smart infrastructure, where precise sensing and secure communication are of utmost importance \cite{zhang2025intelligent}. 

To address these vulnerabilities, Physical Layer Security (PLS) has emerged as a promising solution that capitalizes on the inherent characteristics of the wireless channel—such as fading, interference, and noise. This innovative approach enhances security without relying on conventional cryptographic techniques \cite{10516677, 10003076, devi2025survey}. PLS strategies primarily focus on maximizing secrecy capacity, minimizing eavesdropping risks, and enhancing signal robustness. These objectives are achieved through a variety of techniques, including beamforming, Artificial Noise (AN) generation, and secure waveform design \cite{xu2023beyond, shiu2011physical, abdelsalam2025physical}. Nonetheless, the complexities associated with ISAC—particularly regarding shared spectrum and resource allocation—necessitate the development of more advanced security mechanisms. Such mechanisms must ensure both the reliability of sensing and the confidentiality of communication.
\begin{table}[!t]
\caption{Summary of Important Acronyms}
\centering
\scriptsize 
\begin{tabular}{l|p{0.3\textwidth}}
\toprule
\textbf{Abbreviation} & \textbf{Definition}\\
\midrule
6G & Sixth Generation \\
AI & Artificial Intelligence \\
AN & Artificial Noise \\
AO & Alternate Optimization \\
ARIS & Active RIS \\
BD-RIS & Beyond Diagonal RIS \\
BS & Base Station \\
BER & Bit Error Rate  \\
CRB & Cramér–Rao Bounds \\
CSI & Channel State Information \\
DDPG & Deep Deterministic Policy Gradient \\
DRIS & Diagonal RIS \\
DRL & Deep Reinforcement Learning \\
Eve & Eavesdropper \\
FP & Fractional Programming \\
IoT & Internet of Things \\
ISAC & Integrated Sensing and Communication \\
LoS & Line of Sight \\
ML & Machine Learning \\
MM & Majorization-Minimization  \\
NLoS & Non-Line-of-Sight \\
NOMA & Non-Orthogonal Multiple Access \\
NTNs & Non-Terrestrial Networks \\
PLS & Physical Layer Security \\
RIS & Reconfigurable Intelligent Surfaces \\
RSMA & Rate Splitting Multiple Access \\
S\&C & Sensing and Communication \\
SCA & Successive Convex Approximation \\
SDR & Semidefinite Relaxation \\
SEE & Secrecy Energy Efficiency \\
SNR & Signal to Noise Ratio \\
SOP & Secrecy Outage Probability \\
SR & Secrecy Rate \\
THz & Terahertz \\
UAVs & Unmanned Aerial Vehicles \\
UE & User Equipment \\
V2X & Vehicle-to-Everything \\
WSR & Weighted Sum Rate \\
\bottomrule
\end{tabular}
\label{tab:abbreviation}
\end{table}

A key enabler of PLS in ISAC is Reconfigurable Intelligent Surfaces (RIS), an innovative technology that dynamically manipulates electromagnetic waves to enhance signal transmission, mitigate interference, and improve security \cite{aboagye2022ris,pan2021reconfigurable,basharat2021reconfigurable}. RIS is composed of programmable meta-surfaces that intelligently control wave reflections and transmissions, thereby optimizing the wireless environment. This optimization not only strengthens legitimate communication links but also effectively suppresses potential eavesdropping channels \cite{zhu2023ris, bie2023user, 10833728}. In ISAC networks, RIS serves a dual purpose: it enhances secure communication while simultaneously improving sensing accuracy  and safeguarding the sensing data. Specifically, RIS can direct signals towards intended users, thereby minimizing leakage to unauthorized entities. Additionally, it creates virtual Line-of-Sight (LoS) paths that help circumvent blockages, enhancing jointly the communications and sensing performance in complex environments \cite {10839492, 10720781, 10716670}. These capabilities position RIS as a powerful node for addressing security vulnerabilities and optimizing spectrum efficiency, particularly in 6G-enabled applications that require high security, reliability, and real-time environmental awareness \cite { 10783002}.

This survey underscores the transformative role of RIS in strengthening ISAC PLS. Furthermore, it delves into sophisticated optimization techniques, the technical challenges that arise, and prospective research avenues. This offers valuable insights into the seamless integration of RIS with ISAC security strategies. Such discussions are an essential reference for researchers, engineers, and industry professionals, guiding the progress toward secure, efficient, and scalable ISAC solutions tailored for next-generation 6G networks.

\begin{table*}[ht]
\caption{Survey Papers' Comparison Related to RIS-PLS for ISAC}
\label{table:comparison}
\centering
\renewcommand{\arraystretch}{1.2} 
\begin{tabular}{|c|c|c|c|c|c|c|c|c|}
\hline
\multirow{2}{*}{\textbf{Ref.}} & \multirow{2}{*}{\textbf{Year}} & \multicolumn{4}{c|}{\textbf{Passive RIS-based PLS Schemes}} & \multicolumn{3}{c|}{\textbf{ARIS-based PLS Schemes}} \\
\cline{3-9}
& & \textbf{System Throughput} & \textbf{Cov\&Sen Rates} & \textbf{Sensing SNR} & \textbf{WSR and Sensing} & \textbf{Secrecy Rates} & \textbf{Covert Rates} & \textbf{Sensing SINR} \\
\hline
\cite{cui2021integrating}  & 2021 & $\times$ & $\times$ & $\times$ & $\times$ & $\times$ & $\times$ & $\times$ \\
\hline
\cite{liu2022survey}  & 2022 & $\times$ & $\times$ & $\times$ & $\times$ & $\times$ & $\times$ & $\times$ \\
\hline
\cite{zhou2022integrated}  & 2022 & $\times$ & $\times$ & $\times$ & $\times$ & $\times$ & $\times$ & $\times$ \\
\hline
\cite{zhong2022empowering}  & 2022 & $\times$ & $\times$ & $\times$ & $\times$ & $\times$ & $\times$ & $\times$ \\
\hline
\cite{du2023towards}  & 2023 & $\times$ & $\times$ & $\times$ & $\times$ & $\times$ & $\times$ & $\times$ \\
\hline
\cite{meng2023uav}  & 2023 & * & $\times$ & $\times$ & $\times$ & $\times$ & $\times$ & $\times$ \\
\hline
\cite{wei2023integrated}  & 2023 & $\times$ & $\times$ & $\times$ & $\times$ & $\times$ & $\times$ & $\times$ \\
\hline
\cite{salem2023data}  & 2023 & * & * & $\times$ & $\times$ & $\times$ & $\times$ & $\times$ \\
\hline
\cite{liu20246g}  & 2024 & *** & $\times$ & $\times$ & $\times$ & $\times$ & $\times$ & $\times$ \\
\hline
\cite{gonzalez2024integrated}  & 2024 & *** & $\times$ & ** & $\times$ & $\times$ & $\times$ & $\times$ \\
\hline
\cite{kaushik2024integrated}  & 2024 & * & $\times$ & $\times$ & $\times$ & $\times$ & $\times$ & $\times$ \\
\hline
\cite{magbool2024survey}  & 2024 & * & ** & * & $\times$ & $\times$ & $\times$ & $\times$ \\
\hline
\cite{wen2024survey}  & 2024 & * & * & ** & $\times$ & $\times$ & $\times$ & $\times$ \\
\hline
\cite{wang2024integration}  & 2024 & ** & * & $\times$ & $\times$ & $\times$ & $\times$ & $\times$ \\
\hline
\cite{niu2024interference}  & 2024 & $\times$ & $\times$ & $\times$ & $\times$ & $\times$ & $\times$ & $\times$ \\
\hline
\cite{shtaiwi2024orthogonal}  & 2024 & $\times$ & $\times$ & $\times$ & $\times$ & $\times$ & $\times$ & $\times$ \\
\hline
\cite{qu2024privacy}  & 2024 & ** & ** & *** &$\times$  & $\times$& $\times$ & $\times$ \\
\hline
\cite{jiang2024terahertz}  & 2024 & $\times$ & $\times$ & $\times$ & $\times$ & $\times$ & $\times$ & $\times$ \\
\hline
Our  & 2025 & *** & *** & *** & *** & *** & *** & *** \\
\hline
\multicolumn{9}{l}{\textit{{$\times$} Not Covered, *Preliminary Level, ** Partially Covered, *** Fully Covered}} \\
\end{tabular}
\end{table*}

\subsection{Related Surveys} While ISAC has gained significant attention for its potential to seamlessly integrate sensing and communication, most existing research focuses primarily on the fundamental principles, system architectures, and performance
enhancements, as illustrated in Table \ref{table:comparison}. Despite these advancements, security challenges within ISAC, particularly the role of RIS in enhancing PLS, remain a relatively underexplored area. While some studies acknowledge the vulnerabilities associated with ISAC, they often lack a comprehensive analysis of how RIS can effectively address emerging threats, such as eavesdropping, jamming, and spoofing.\\
Numerous studies, including \cite{cui2021integrating},\cite{liu2022survey},\cite{zhou2022integrated},\cite{zhong2022empowering},\cite{du2023towards},\cite{wei2023integrated},\cite{niu2024interference}, \cite{shtaiwi2024orthogonal}, \cite{jiang2024terahertz} 
 have examined the core principles of ISAC, emphasizing its potential to integrate wireless communication with radar sensing. These studies largely focus on ISAC system design, signal processing methodologies, and performance trade-offs. Nevertheless, they predominantly neglect the importance of RIS in augmenting ISAC security. The study in \cite{cui2021integrating} classifies the predominant ISAC solutions according to the layers in which integration occurs. The research in \cite{liu2022survey} summarizes the main performance measurements and constraints utilized in ISAC. The research in \cite{zhou2022integrated} provides a thorough examination of the cutting-edge methodologies pertaining to the ISAC methodology, focusing on waveform design. The study in \cite{zhong2022empowering} examines cutting-edge enabling technologies by analyzing current advancements in ISAC-assisted beamforming technology within vehicle networks. The study in \cite{du2023towards} provides a thorough examination of the implementation of ISAC algorithms in Vehicle-to-Infrastructure (V2I) networks. The study in \cite{wei2023integrated} provides an analysis of the literature on ISAC signals in the context of mobile communication systems, including ISAC signal design, processing, and optimization. The article \cite{niu2024interference} presents a thorough examination of interference management approaches in ISAC systems, encompassing network architecture, system design, signal processing, and resource allocation. The research in  \cite{shtaiwi2024orthogonal} provides a complete examination of recent research advancements and the current state of orthogonal time frequency space (OTFS)-assisted ISAC systems to attain a full understanding of the current landscape and progress. In \cite{jiang2024terahertz} intricate design of sensing and communication systems is examined thoroughly. Practical trials, demonstrations, and experiments are summarized as well.
 
 Several studies including, \cite{meng2023uav}, \cite{salem2023data}, \cite{liu20246g},\cite{gonzalez2024integrated}, \cite{kaushik2024integrated}\cite{magbool2024survey}, \cite{wen2024survey},\cite{wang2024integration}, \cite{qu2024privacy} have investigated the function of passive RIS in augmenting security for ISAC systems. These studies largely emphasize the advantages of RIS, such as enhanced signal strength and interference mitigation, in safeguarding ISAC transmissions. The majority focus primarily on system throughput and enhancements in overall performance. These studies offer abstract discussions of the possibility of RIS for protecting ISAC but fail to give a comprehensive study of essential security characteristics, including secrecy, integrity, authentication, and resilience against adversarial attacks. Research in \cite{meng2023uav} explores UAV motion control, wireless resource allocation, and interference management in ISAC systems utilizing both single and multiple UAVs. The study \cite{salem2023data} aims to examine the integration of several ML approaches into ISAC systems, addressing multiple applications. In \cite{liu20246g}, a novel hybrid approach for ISAC channel modeling is described, wherein the modeling process is divided into three components: targets, clutter, and interference. The authors in \cite{gonzalez2024integrated} present a vision for ISAC networks and an outline of how signal processing, optimization, and ML approaches might be utilized to actualize them within the framework of 6G. The research in \cite{kaushik2024integrated} discusses some particulars of ISAC from an IoT standpoint in 6G, facilitating diverse contemporary IoT applications and essential technological enablers. In \cite{magbool2024survey}, multiple distinctive characteristics of ISAC have been discussed from an IoT perspective in 6G, facilitating diverse contemporary IoT applications and essential technological enablers. The authors in \cite{wen2024survey} subsequently examines the cutting-edge signal designs for ISAC, in addition to network resource management methodologies. The study \cite{wang2024integration} examines solutions or metaverse scenarios, including smart homes, smart factories, smart healthcare, smart transportation, and UAV networks, focusing on recognition accuracy,  communication latency, user data privacy, and communication dependability. Finally, \cite{qu2024privacy} outlines the technical impediments that hinder the establishment of a secure and private ISAC environment within traditional network architectures. Despite ongoing advancements, there remains a significant research gap in understanding how RIS can be strategically leveraged to improve the security of ISAC systems. Addressing this gap requires a holistic approach that evaluates both passive and Active RIS (ARIS) configurations, their effectiveness in mitigating emerging security threats, and their practical deployment considerations in future wireless networks.\\
\vspace{-22pt}
\subsection{Motivation and Contribution}The convergence of RIS and ISAC in 6G networks presents both challenges and opportunities for enhancing PLS. In ISAC environments, traditional security mechanisms frequently face difficulties in effectively balancing the intricate trade-offs among sensing, communication, and secrecy constraints. RIS technology, recognized for its ability to dynamically reconfigure wireless propagation, emerges as a promising strategy to enhance both security and efficiency within these complex scenarios. However, a notable gap exists in our understanding of how varying RIS configurations, modes, and architectural designs influence the security framework within ISAC systems. This survey seeks to bridge this knowledge gap by providing a comprehensive and structured analysis of both passive RIS and ARIS techniques. We will elucidate their respective advantages, limitations, and critical deployment considerations that are vital for strengthening ISAC-enabled 6G networks.
By thoroughly addressing these vital aspects, this survey substantially enhances the understanding of RIS-enhanced PLS within ISAC systems.

\begin{itemize} 
\item This survey establishes a strong foundation by covering the fundamentals of RIS, PLS, and ISAC, providing a structured understanding of their integration in 6G networks. It explores RIS technology, detailing Diagonal (D)-RIS and Beyond Diagonal (BD)-RIS architectures, their operating modes (reflective, transmissive, hybrid, and multi-sector), and their advantages and limitations in ISAC. The paper also examines PLS technologies, including channel-based, key-based, and signal processing-based PLS.
\item The survey categorizes RIS-assisted PLS approaches into passive RIS-based and ARIS-based techniques. It systematically reviews how passive RIS enhances ISAC security by optimizing system throughput, covert and Secrecy Rates (SRs), sensing SNR, and Weighted Sum Rate (WSR) under various constraints. For ARIS-based approaches, it highlights how active beamforming further improves secrecy and covert rates while maintaining robust sensing performance under communication and security constraints.

\item Beyond classification, this work provides a comparative analysis of RIS-PLS techniques, evaluating trade-offs between security, sensing, and communication efficiency while addressing practical challenges such as optimization complexity, scalability, and real-world deployment constraints. Lastly, the survey identifies key open research challenges and proposes future directions, offering novel optimization strategies to enable secure, efficient, and scalable RIS-PLS solutions for next-generation ISAC networks.

\end{itemize}

\begin{figure}
\centering
\includegraphics[width=0.40\textwidth, height=0.70\textheight]{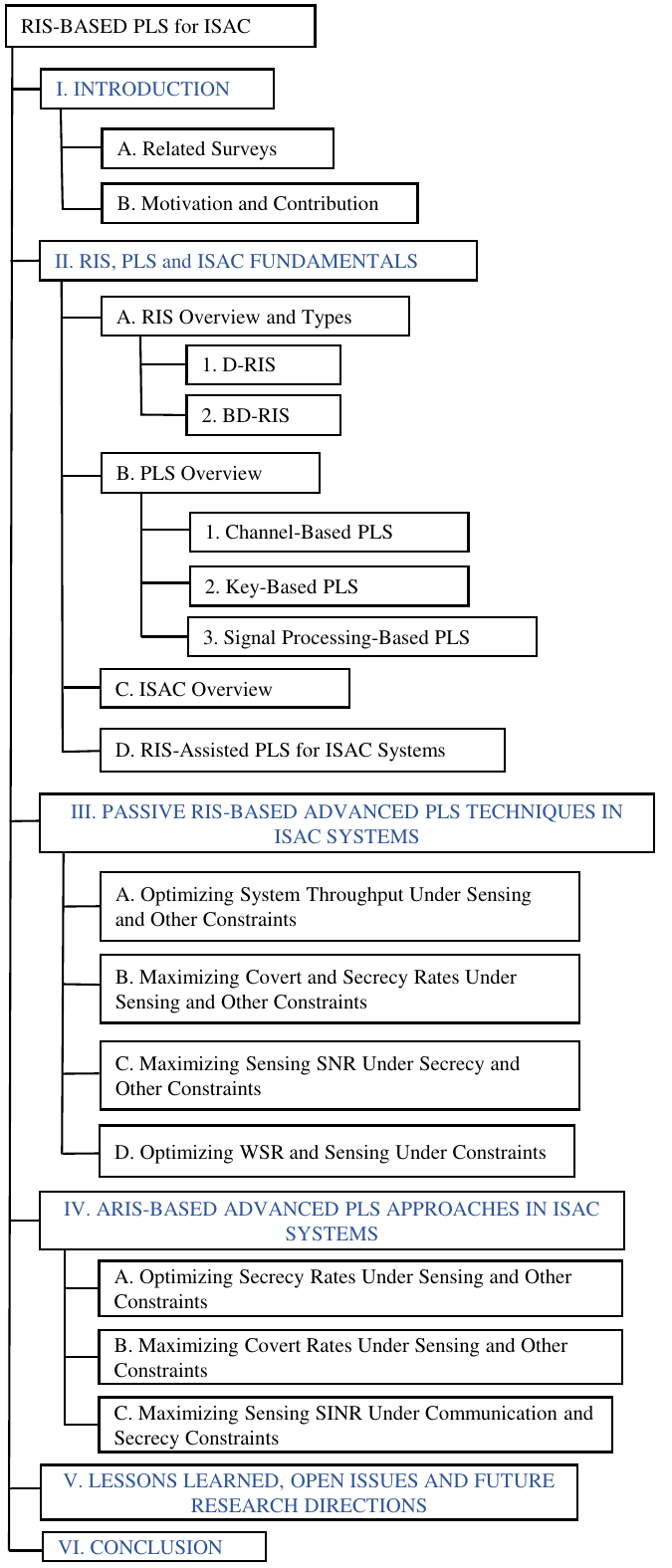}
\caption{Survey Taxonomy}
\label{fig:Taxonomy}
\end{figure}

This survey explores RIS-based PLS for ISAC in a structured manner. Section \ref{Sec: sectionII} delves into the fundamentals of RIS technology, including the distinctions between D-RIS and BD-RIS. It presents an in-depth analysis of their operational modes, architectures, and configurations, establishing a solid foundation for understanding their applications. Furthermore, this section provides a comprehensive overview of PLS, elucidating its critical role in enhancing the security of ISAC. It explores various sensing approaches and highlights the advantages of RIS-assisted PLS, laying the groundwork for subsequent discussions.
Section \ref{sec:sectionIII} concentrates on passive RIS-based PLS techniques. Transitioning smoothly, Section \ref{sec:sectionIV} delves into ARIS-based PLS strategies, emphasizing the significance of active beamforming. It illustrates how this approach enhances secrecy, facilitates covert communication, and improves sensing performance within ISAC frameworks. In Section \ref{sec:sectionV}, valuable lessons are drawn from the discussions, identifying research challenges and proposing future directions. This section pinpoints key areas ripe for further exploration and innovation in the realm of RIS-based PLS for ISAC, encouraging continued advancement in this field.
Finally, Section \ref{sec:sectionVI} encapsulates the key findings of the survey and offers a comprehensive conclusion on RIS-based PLS within the context of ISAC. A visual taxonomy is provided in Fig. \ref{fig:Taxonomy}
\vspace{-15pt}

\section{RIS, PLS, and ISAC Fundamentals}
\label{Sec: sectionII}
The section provided a comprehensive overview of the fundamentals of RIS, PLS, and ISAC. It introduced the essential concepts of RIS, specifically discussing two types: D-RIS, with an examination of its advantages and limitations, and BD-RIS, which focused on architectural configurations, operating modes, and their respective pros and cons. The subsection on PLS highlighted three primary categories: channel-based PLS techniques aimed at maximizing secrecy capacity, key-based PLS focusing on physical layer key generation, and signal processing-based PLS that targets secure transmission methods. Additionally, the fundamentals of ISAC were briefly discussed, outlining the different types of ISAC sensing. Finally, the integration of RIS-assisted PLS within ISAC systems was addressed, describing the advantages of such approaches. 
\vspace{-15pt}
\subsection{RIS Overview and Types:}
RIS technology is revolutionizing wireless communication by enabling intelligent control over the electromagnetic wave propagation. Unlike traditional passive environments, RIS facilitates dynamic and adaptive wireless configurations, featuring an array of reconfigurable elements that adjust both the amplitude and phase of signals \cite{8930608}\cite{ahmed2024survey}. This remarkable flexibility significantly enhances signal strength, minimizes interference, and improves coverage, particularly in challenging environments where conventional methods may struggle. Moreover, RIS plays a pivotal role in next-generation systems like 6G, where the emphasis is on Energy Efficiency (EE), spectral efficiency, and ultra-reliable communication \cite{10776998}. The technology offers multiple noteworthy advantages, with EE being a primary benefit. Operating with minimal power consumption, RIS utilizes passive elements that require only limited energy for control \cite{8741198}. This characteristic positions RIS as a more sustainable alternative to traditional relay systems, which heavily depend on power amplifiers. Additionally, this capability facilitates precise control of wave propagation and enables the implementation of advanced beamforming techniques \cite{10032238}. RIS can be categorized as passive, active, or hybrid, each serving different roles in enhancing wireless communications.

Passive RIS relies on metasurfaces to reflect signals without amplification, thereby enhancing coverage, mitigating interference, and improving security, all while maintaining EE \cite{zhang2022active, schroeder2021passive}. However, its reliance on base stations (BSs) or user equipment (UEs) for channel estimation presents challenges, particularly in dynamic environments where it is vulnerable to double fading \cite{di2022communication, 9860805, najafi2020physics}. To counter these limitations, ARIS introduces the use of active-load impedances, phase shifters, and amplifiers. These components work together to amplify signals, thereby enhancing beamforming capabilities, broadening coverage, and improving interference suppression \cite{jian2022reconfigurable, lv2022ris}. ARIS effectively mitigates double fading, ensuring stronger signal propagation while achieving higher SNR and efficiency with fewer elements \cite{guo2023enhanced, najafi2020physics, niu2022joint}. Moreover, its ability to seamlessly switch between active and passive modes enhances operational flexibility, making it particularly advantageous for energy-constrained deployments \cite{rao2023active, wang2024beamforming}. In pursuit of a balance between performance and efficiency, Hybrid RIS (HRIS) combines passive reflection with active amplification. This innovative approach improves beamforming and power control, leading to enhanced adaptability in next-generation networks \cite{yildirim2021hybrid, 10584518, he2021channel}. While passive RIS prioritizes EE, ARIS places a greater emphasis on performance through increased power consumption. In contrast, HRIS offers a scalable and adaptive solution that meets the evolving needs of wireless communication.

\subsubsection{D-RIS  Overview}

D-RIS leverages a reconfigurable matrix featuring only diagonal phase-shifts response, allowing each element to independently modulate the amplitude and phase of the incoming signals. However, this configuration imposes limitations on its beamforming capabilities. To address this, each element features a tunable meta-atom that allows for precise phase adjustments, which play a crucial role in steering signals towards their intended destination. This enhancement not only boosts signal strength but also reduces interference. D-RIS operates in three distinct modes: reflective, transmissive, and hybrid (STAR), with each mode tailored for specific scenarios. A detailed discussion of these modes follows below.

\begin{figure}
\centering
\includegraphics[width=0.43\textwidth]{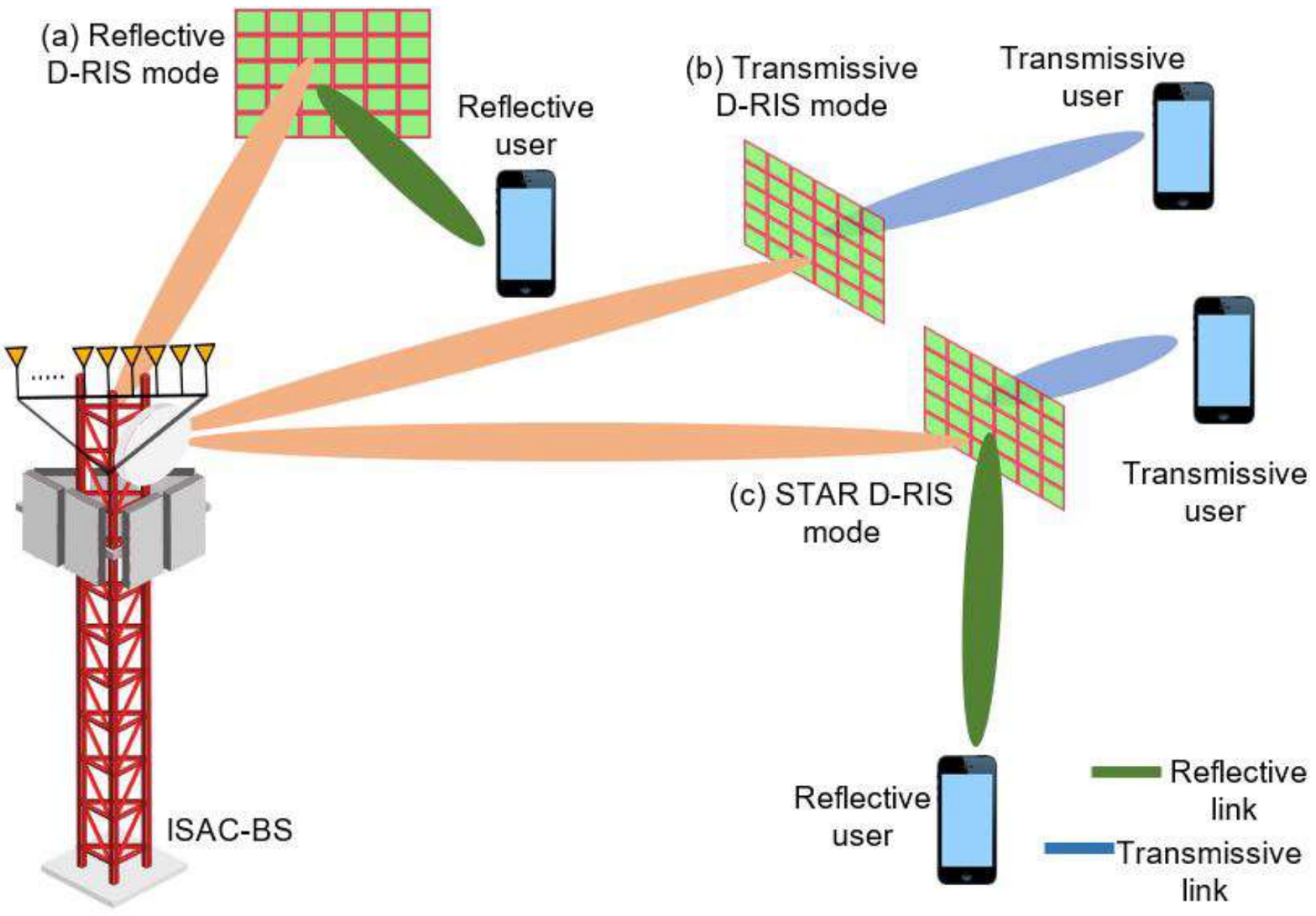}
\caption{Operational modes of D-RIS: (a) Reﬂective D-RIS, (b) Transmissive D-RIS, and (c) Hybrid D-RIS}
\label{fig:operational modes}
\end{figure}

\paragraph{Reflective Mode}
In reflective mode, RIS elements effectively reflect the incoming signals toward the receiver when both the transmitter and receiver are positioned on the same side of the surface, as shown in Fig. \ref{fig:operational modes}a. The phase shift matrix is pivotal, enabling the independent adjustment of each element's phase, which facilitates the precise steering of the reflected signals. This capability significantly enhances coverage and signal quality. Nonetheless, a significant limitation of this mode is the risk of self-interference, as the incident and reflected signals coexist on the same side of the surface. \cite{8796365}

\paragraph{Transmissive Mode}
In transmissive mode, the RIS elements facilitate the passage of signals while concurrently manipulating their phase, as shown in Fig. \ref{fig:operational modes}b.  This configuration proves advantageous when the RIS is situated between the transmitter and receiver or when it functions as a transmitter itself. Furthermore, it offers an energy-efficient alternative to conventional antenna arrays \cite{khan2024cr}.

\paragraph{STAR-RIS}
 It enables each element to simultaneously reflect and transmit signals, ensuring continuous coverage on both sides of the surface, as shown in Fig. \ref{fig:operational modes}c. Moreover, this mode enhances flexibility by facilitating seamless communication, as it dynamically adjusts the reflection and transmission properties for optimal performance. \cite{9570143, 10133841}


\subsubsection*{\textbf{Advantages and Limitations of D-RIS}}

D-RIS offers several advantages that make it a key technology for 6G networks. Its simple operational principle, where each element independently adjusts the phase shift, simplifies hardware design and system implementation. The reflective matrix remains diagonal, making it easier to model and control for specific applications \cite{10720781}. Additionally, D-RIS is highly energy-efficient, as it operates passively without requiring power-hungry components like amplifiers or RF chains. This minimal power requirement makes it a more energy-efficient alternative to traditional active relays \cite{ahmed2025advancements}. D-RIS is also versatile, capable of enhancing both LoS and Non-Line-of-Sight (NLoS) communication, improving signal coverage in diverse environments such as urban, rural, indoor, and non-terrestrial setups.
\setlength{\textfloatsep}{5pt}
However, D-RIS has several limitations. One significant drawback is its lack of inter-element communication, which hampers its ability to precisely control wavefronts. This limitation diminishes its effectiveness in complex environments, particularly in scenarios characterized by dense multipath or interference \cite{10715713}. Moreover, the independent control of elements can create performance bottlenecks. This is especially evident when managing multi-user interference or when fine-grained beam steering is required. Additionally, D-RIS struggles to adapt to real-time changes in dynamic environments, such as user mobility or fluctuating obstacles. Consequently, it is less suited for dynamic networks, such as those used in vehicular or drone-assisted communications \cite{10833728}. These limitations have spurred the exploration of advanced RIS technologies, like BD-RIS, to overcome these constraints and provide more dynamic, adaptable signal manipulation.

\subsubsection{BD-RIS}

The BD-RIS represents a significant advancement over traditional D-RIS technology by integrating non-diagonal components into its phase shift matrix \cite{9514409}. This enhancement empowers BD-RIS to precisely manage electromagnetic waves, optimizing beam patterns, expanding network coverage, and effectively minimizing interference \cite{9913356}. Moreover, BD-RIS can dynamically adjust to environmental changes and varying user requirements, while still preserving conventional operational modes—reflective, transmissive, and hybrid. In addition, it introduces an innovative multi-sector mode. This new mode further enhances system performance through the utilization of high-gain, narrow-beamwidth elements, ensuring comprehensive spatial coverage.

BD-RIS architecture fundamentally differs from D-RIS due to its interconnected structure, which facilitates superior beamforming capabilities. In BD-RIS, the phase shift matrix is fully connected, in contrast to D-RIS, which exclusively utilizes diagonal elements \cite{10159457,del2024physics}. The BD-RIS phase shift matrix satisfies the following constraint:
\begin{equation}
\boldsymbol{\Phi}_{\text{BD-RIS}} \boldsymbol{\Phi}_{\text{BD-RIS}}^H = \textbf{I}_N, \quad N = 4,
\label{eq1}
\end{equation}

where \(\boldsymbol{\Phi}_{\text{BD-RIS}}\) is the complex phase shift matrix, \(\boldsymbol{\Phi}_{\text{BD-RIS}}^H\) denotes the Hermitian transpose of \(\boldsymbol{\Phi}_{\text{BD-RIS}}\), and \(\textbf{I}_N\) represents an \(N \times N\) identity matrix. Conversely, the constraint for D-RIS phase shift elements is given by:
\begin{equation}
|\phi_{n,n}|^2 = 1, \quad \forall n \in \{1, 2, 3, 4\},
\label{eq2}
\end{equation}
where \(\phi_{n,n}\) represents the individual diagonal elements of the phase shift matrix.

\subsubsection*{\textbf{BD-RIS Architectural Configurations}}

BD-RIS architectures can be classified into three main configurations: single-connected, fully-connected, and group-connected, each with distinct complexity and performance trade-offs \cite{10716670}.

\paragraph{Single-connected}

The simplest configuration of a single-connected BD-RIS consists of independently functioning elements that operate without any interconnections \cite{10316535, khan2024integrationUAV}. The phase shift matrices in this scenario are diagonal:
\begin{equation}
{\bf \Phi}_r = \text{diag}(\phi_{r,1}, \phi_{r,2}, \phi_{r,3}, \phi_{r,4}),
\label{eq3}
\end{equation}
\begin{equation}
{\bf \Phi}_t = \text{diag}(\phi_{t,1}, \phi_{t,2}, \phi_{t,3}, \phi_{t,4}),
\label{eq4}
\end{equation}
subject to energy conservation constraints:
\begin{equation}
|\phi_{r,n}|^2 + |\phi_{t,n}|^2 = 1, \quad \forall n,
\label{eq5}
\end{equation}
where \(\phi_{r,n}\) and \(\phi_{t,n}\) represent reflection and transmission coefficients, respectively, for the \(n\)th element.

\paragraph{Fully-connected}

The fully-connected BD-RIS configuration interconnects all elements through an impedance network, allowing for non-diagonal matrices that significantly enhance beamforming flexibility \cite{10716670}. The phase shift matrices must fulfill a unitary condition:
\begin{equation}
{\bf \Phi}_r^H {\bf \Phi}_r + {\bf \Phi}_t^H {\bf \Phi}_t = {\bf I}_N,
\label{eq6}
\end{equation}
where \({\bf \Phi}_r\) and \({\bf \Phi}_t\) represent the reflection and transmission phase shift matrices, respectively.

\paragraph{Group-connected}

Group-connected BD-RIS organizes elements into multiple fully interconnected subarray clusters, effectively managing scalability \cite{9913356}. Its phase shift matrices adopt a block-diagonal structure:
\begin{equation}
{\bf \Phi}_r = \text{blkdiag}({\bf \Phi}_{r,1}, {\bf \Phi}_{r,2}),
\label{eq7}
\end{equation}
with subarray matrices meeting the following condition:
\begin{equation}
\boldsymbol{\Phi}_{r,g}^H \boldsymbol{\Phi}_{r,g} = {\bf I}_{\bar{N}},
\label{eq8}
\end{equation}
where \(\boldsymbol{\Phi}_{r,g}\) is the reflection phase shift matrix for each group and \(\bar{N}\) is the number of elements per group. This configuration balances performance and complexity, ideal for large-scale deployments.

\subsubsection*{\textbf{Operating Modes of BD-RIS}}

BD-RIS operates effectively in four distinct modes: reflective, transmissive, hybrid, and multi-sector, each tailored for specific communication conditions.

\paragraph{Reflective Mode}

In reflective mode, the signals reflect toward the receiver with the transmitter and receiver positioned on the same side \cite{10158988}. The phase shift matrix satisfies:
\begin{equation}
\boldsymbol{\Phi}_r^H \boldsymbol{\Phi}_r = {\bf I}_N.
\label{eq9}
\end{equation}
This mode is optimal for half-space coverage scenarios.

\paragraph{Transmissive Mode}

In transmissive mode, signals traverse through BD-RIS, enhancing obstacle penetration \cite{10839492}. The phase shift matrix has block-diagonal constraints:
\begin{equation}
\boldsymbol{\Phi}_t = \text{blkdiag}(\boldsymbol{\Phi}_{t,1}, \boldsymbol{\Phi}_{t,2}),
\label{eq10}
\end{equation}
subject to:
\begin{equation}
\boldsymbol{\Phi}_{t,1}^H \boldsymbol{\Phi}_{t,1} = {\bf I}_4, \quad \boldsymbol{\Phi}_{t,2}^H \boldsymbol{\Phi}_{t,2} = {\bf I}_4.
\label{eq11}
\end{equation}
\paragraph{STAR-RIS}
The STAR mode seamlessly integrates both reflective and transmissive operations, enabling dual-sided communication. This innovative approach enhances flexibility by dynamically adjusting both reflection and transmission, ultimately improving overall performance \cite{10817342}.
\paragraph{Multi-sector Mode}
The multi-sector mode effectively partitions the BD-RIS surface into independent sectors, allowing for targeted spatial coverage. This mode utilizes high-gain, narrow beamwidth elements, which significantly enhance precision in both beamforming and overall spatial coverage \cite{10316535}.
\begin{figure}
\centering
\includegraphics[width=0.44\textwidth]{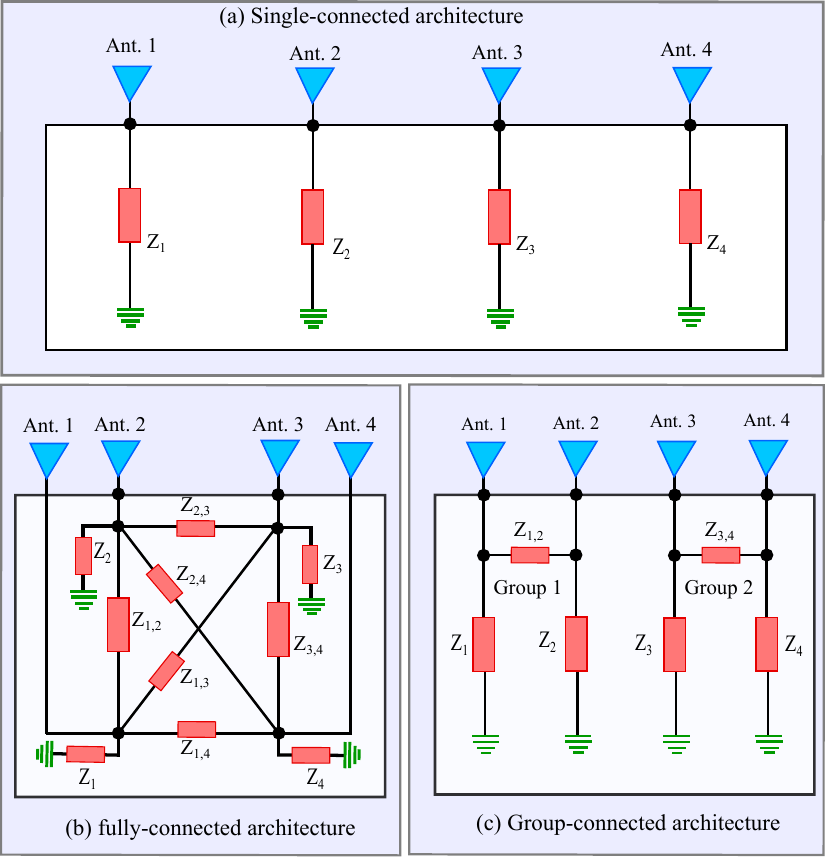}
\caption{BD-RIS architectures: (a) Single connected, (b) Fully-connected, (c) Group-connected.}
\label{fig:BDRISarchitecture}
\end{figure}

\subsubsection*{\textbf{Advantages and Disadvantages of BD-RIS}}

BD-RIS presents significant advantages over traditional D-RIS, positioning it as a highly promising technology for 6G networks. The primary benefit of BD-RIS lies in its advanced beamforming capabilities, facilitated by its interconnected elements. This technological innovation enables the simultaneous control of both phase and amplitude, leading to a considerable enhancement in signal strength, quality, and flexibility across various directions and diverse communication scenarios \cite{9913356}. Additionally, this work highlights that BD-RIS demonstrates improved adaptability to dynamic environments. Consequently, BD-RIS ensures reliable performance, making it particularly well-suited for dynamic urban settings or vehicular networks \cite{9514409}. 

Another significant advantage of BD-RIS is its effective management of interference. By adjusting both amplitude and phase, BD-RIS significantly reduces interference in a more comprehensive manner than D-RIS \cite{10316535}. Moreover, BD-RIS provides complete 360-degree coverage, which surpasses the typical 180-degree coverage offered by D-RIS. This capability is particularly advantageous for ensuring uniform signal distribution in extensive applications, such as industrial IoT and Non-Terrestrial Networks (NTNs). As a result, BD-RIS emerges as a strong contender for future communication frameworks, delivering versatile and reliable solutions well-suited for a variety of implementation scenarios. For a detailed comparison of different RIS types, please refer to Table \ref{tab:DRIS_vs_BDRIS}.

\begin{table}[htbp]
\centering
\renewcommand{\arraystretch}{1.2} 
\setlength{\tabcolsep}{3pt} 
\caption{Comparison of D-RIS and BD-RIS}
\label{tab:DRIS_vs_BDRIS}

\begin{tabular}{|p{2cm}|p{3cm}|p{3cm}|}  
\hline
\textbf{Feature} & \textbf{D-RIS } & \textbf{BD-RIS } \\
\hline
\textbf{Scattering Model} & Independent elements (diagonal) & Inter-element coupling (non-diagonal) \\
\hline
\textbf{Signal Control} & Phase shifting only & Phase shifting + amplitude control \\
\hline
\textbf{Complexity} & Low & Moderate to High \\
\hline
\textbf{Hardware Design} & Nearly Passive or Active & Nearly Passive or Active \\
\hline
\textbf{Energy Consumption} & Low (no external power required) & Higher (if active elements are used) \\
\hline
\textbf{Performance} & Basic beamforming & Advanced wave manipulation \\
\hline
\textbf{Applications} & Reflection control, basic beamforming & Secure communication, interference mitigation \\
\hline
\end{tabular}
\end{table}
\vspace{-8pt}
\subsection{PLS Overview}
In this subsection, PLS emerges as a contemporary strategy for securing wireless communications. PLS leverages the unique characteristics inherent in wireless channels—such as noise, fading, and interference—thus moving beyond the traditional reliance on cryptographic techniques. By exploiting these natural imperfections, PLS significantly enhances security by facilitating reliable decoding for legitimate users, while simultaneously obstructing eavesdroppers' (Eves) attempts to intercept transmissions \cite{hamamreh2018classifications}. A fundamental concept within PLS is the notion of secrecy capacity, which defines the maximum rate at which confidential information can be securely transmitted, assuming that the legitimate receiver enjoys a superior channel compared to that of the Eve \cite{kapetanovic2015physical, mukherjee2011full, xu2017proactive, xu2015proactive}.  



PLS also addresses a variety of threat models. One such model involves interceptive attacks, where passive eavesdroppers, referred to as Eves, monitor signals without direct interference. Additionally, proactive adversaries may manipulate Channel State Information (CSI) to gain an advantage, further complicating the security landscape. Another significant threat comes from jamming-based attacks. In these cases, intentional interference is introduced to disrupt communications. Interestingly, in some scenarios, this jamming can also enhance the Eve's ability to receive signals, posing a dual threat. Moreover, identity spoofing presents further risks, as attackers may impersonate legitimate users to gain unauthorized access, exploiting the inherently open nature of wireless channels \cite{xiao2009channel}.  

To counter these risks, physical layer-key generation models utilize randomness extracted from CSI, Received Signal Strength (RSS), and phase information to establish secure keys through reconciliation and privacy amplification techniques \cite{bennett1995generalized}. Additionally, physical layer authentication mechanisms are essential for verifying that transmitted signals originate from legitimate users. This is accomplished by analyzing unique CSI patterns and employing hypothesis testing \cite{xiao2007fingerprints, xiao2008using, liu2013two}. By leveraging spatial variations in CSI and coherence properties, these methods effectively authenticate devices, thereby preventing unauthorized access. This capability capitalizes on the dynamic nature of wireless channels, which are continuously influenced by evolving communication technologies. Notably, the potential of PLS to harness the inherent characteristics of wireless channels, instead of relying solely on traditional cryptographic encryption, positions it as an innovative solution for secure and efficient ISAC systems. 

\subsubsection{Channel-Based PLS (Secrecy Capacity Techniques)}
This approach secures wireless communication by exploiting the unique characteristics of the physical channel between legitimate users and potential Eves. The fundamental idea is to ensure that the legitimate receiver experiences superior channel conditions compared to an unauthorized entity, thereby maximizing the SR. Key techniques in this category include:

\begin{table*}[htbp]
\centering
\caption{Comparison of PLS Techniques }
\label{table:PLS_Comparison}
\renewcommand{\arraystretch}{1.3} 
\setlength{\tabcolsep}{5pt} 
\resizebox{0.95\textwidth}{!}{ 
\begin{tabular}{|p{2.8cm}|p{3.5cm}|p{4.5cm}|p{2.2cm}|p{2.2cm}|p{3.5cm}|}
\hline
\textbf{PLS Type} & \textbf{Security Principle} & \textbf{Techniques} & \textbf{Complexity} & \textbf{Cost Effectiveness} & \textbf{Hardware Requirement} \\ 
\hline

Channel-Based PLS  
& Exploits physical channel differences  
& Wiretap Channel Model, AN, Beamforming, Cooperative Jamming  
& Moderate to High  
& High  
& Low (Only signal processing) \\ 
\hline

Key-Based PLS  
& Uses channel randomness for encryption  
& Reciprocity-Based Key Generation, Random Modulation, Pilot Contamination Defense  
& Moderate  
& Moderate  
& Low (Channel measurement hardware) \\ 
\hline

Signal Processing PLS  
& Embeds security in signal design  
& Directional Modulation, Secret Key Embedding, Covert Communication  
& High  
& Moderate to Low  
& Moderate (Advanced modulation equipment) \\ 
\hline

\end{tabular}
} 
\end{table*}

\begin{itemize}
    \item Wiretap Channel Model: Based on Shannon’s information-theoretic security principles, this technique ensures that the intended user receives a higher channel capacity than the Eve, thereby minimizing information leakage \cite{ghadi2024physical}.
\item AN Injection: Introduces controlled noise into the transmission in directions where potential Eves are located while maintaining clear communication for the legitimate receiver \cite{cheng2025artificial}.
\item Beamforming: Directs signal power toward the intended recipient while minimizing energy leakage toward unintended users, thereby enhancing security \cite{liu2013two}.
\item Cooperative Jamming: Involves the use of friendly nodes that generate interference to degrade the Eve’s ability to decode the message while keeping the communication channel clear for the legitimate user \cite{chi2024performance}. For instance, in a wireless network, a BS can employ beamforming and AN to ensure that only the intended user receives an intelligible signal while preventing adversaries from intercepting the transmission.
\end{itemize}

\subsubsection{Key-Based PLS (Physical Layer Key Generation)} 
Unlike traditional cryptographic approaches that rely on pre-shared keys, key-based PLS leverages the random and time-varying characteristics of the wireless channel to generate secret encryption keys dynamically  \cite{wang2020power, xiao2007fingerprints, xiao2008using, liu2013two}. Since each communication link experiences unique multipath fading, Eve cannot easily replicate the key. Common techniques include:
\begin{itemize}

\begin{table*}[htbp]
\centering
\renewcommand{\arraystretch}{1.2}
\caption{Comparison of ISAC Sensing Types}
\label{tab:ISAC_Sensing_Types}
\begin{tabular}{|p{3.2cm}|p{4.5cm}|p{4.5cm}|p{4.5cm}|}
\hline
\textbf{Sensing Type} & \textbf{Definition} & \textbf{Advantages} & \textbf{Challenges} \\ 
\hline
\textbf{Monostatic Sensing} & Transmitter and receiver are colocated & Simple implementation, easy integration & Limited detection range, affected by environmental clutter \\ 
\hline
\textbf{Bistatic Sensing} & Transmitter and receiver are positioned separately & Wider coverage, better spatial diversity & Requires synchronization, complex signal processing \\ 
\hline
\textbf{Multistatic Sensing} & Multiple transmitters and receivers work together & High spatial resolution, robust target detection & Coordination complexity, increased computational demands \\ 
\hline
\end{tabular}
\end{table*}

\item Reciprocity-Based Key Generation: Utilizes the principle that the wireless channel between two communicating parties exhibits symmetrical properties, allowing them to extract identical secret keys based on channel variations \cite{waqas2018social}.
\item Random Modulation: Applies intentional signal distortions known only to the legitimate receiver, making it difficult for an adversary to decipher the communication.
\item  Pilot Contamination Defense: Alters pilot signals used for channel estimation to prevent Eves from accurately estimating the secret key \cite{elijah2015comprehensive}. For example, in a WiFi network, two devices can measure random fluctuations in the signal’s fading pattern and generate a shared encryption key, which remains unknown to an external Eve.
\end{itemize}
\subsubsection{Signal Processing-Based PLS} 
This category enhances communication security by embedding security measures directly into the transmitted signals \cite{huang2025joint}. It prevents unauthorized access through advanced signal processing and transmission techniques designed to obscure information from Eves while maintaining signal clarity for intended users \cite{ansari2022directional}. Key methods include:
\begin{itemize}
    
\item Directional Modulation (DM): Alters the phase and amplitude of the signal dynamically, ensuring that the message transmitted is intelligible only at a specific angle, rendering it undecodable elsewhere \cite{masoud2025robust}.
\item Secret Key Embedding: Integrates encryption keys within the physical signal waveform itself, making it accessible only to authorized users with the appropriate decoding mechanism.
\item Covert Communication: Conceals transmission by embedding signals within ambient noise or legitimate background interference, making detection and interception by Eves highly challenging \cite{duan2025adaptive}.
A practical example is a satellite communication system that transmits information in a manner that ensures only receivers positioned at a precise location can decode the message while preventing interception by unauthorized entities. Comparison of different PLS types is given in Table \ref{table:PLS_Comparison}.
 
\end{itemize}

\subsection{ISAC Overview}

ISAC represents the convergence of wireless communication technologies, including wireless networks and signal processing, with radar-based sensing techniques into a unified framework, leveraging wireless RF signals to simultaneously perform communication and environmental sensing tasks \cite{kaushik2024integrated, wang2024integration}. By integrating these traditionally separate domains, ISAC achieves synchronized multimodal data acquisition and transmission, enhancing system efficiency and performance.

ISAC capitalizes on shared resources—such as frequency bands, hardware components, and computational platforms—to build systems capable of simultaneous sensing and communication tasks \cite{zhou2022integrated, lu2024integrated}. In practice, this integration transforms communication networks into active sensory systems. Specifically, network infrastructure uses emitted radio signals to detect, track, and characterize objects within its operational environment. The sensing capability includes acquiring detailed information such as range, velocity, position, orientation, shape, and even material composition, thus enabling intelligent environmental interactions \cite{kaushik2024integrated}.

The integration of sensing data into communication processes significantly improves network performance. For instance, CSI and localization data collected through sensing enhance core communication tasks like channel estimation, adaptive beamforming, and interference management \cite{zhou2022integrated, wang2024integration}. This adaptive capability is crucial for maintaining reliable and efficient communication in highly dynamic environments, such as dense urban areas, where obstacles and user mobility frequently alter signal propagation conditions \cite{wang2024integration, volgushev2024integrated}.
Conversely, communication capabilities enhance sensing functions by providing timely data exchange and feedback, thereby improving accuracy and robustness in object detection, classification, and localization. Effective coordination among sensing nodes further reduces redundancy and optimizes resource utilization. This mutual enhancement makes ISAC exceptionally valuable for critical applications including emergency response, surveillance, environmental monitoring, and military operations, where precision and timely data collection are paramount \cite{wang2024integration,  kaushik2024integrated}.

\subsubsection*{\textbf{Types of ISAC Sensing}}

ISAC systems utilize three primary types of sensing: monostatic, bistatic, and multistatic. Each type offers distinct benefits and suits specific application scenarios.

\paragraph{Monostatic Sensing}
In monostatic sensing, a single device performs both signal transmission and reception. This method simplifies the system architecture, making it particularly appropriate for compact or power-constrained environments. Monostatic sensing systems analyze the transmitted signal reflections, measuring parameters such as distance and velocity via time delays and Doppler shifts \cite{volgushev2024integrated}.

\paragraph{Bistatic Sensing}
Bistatic sensing involves separate devices for signal transmission and reception, allowing flexible deployment and coverage from multiple observation angles. This spatial flexibility facilitates enhanced tracking of moving objects and detailed terrain mapping, proving beneficial in complex scenarios where multiple perspectives improve detection and tracking reliability \cite{wang2024integration}.

\paragraph{Multistatic Sensing}
Multistatic sensing employs multiple transmitters and receivers strategically distributed across diverse locations. This configuration significantly enhances spatial coverage, accuracy, and reliability. Multistatic systems excel in large-scale applications, providing comprehensive monitoring capabilities from multiple vantage points, ideal for scenarios like extensive surveillance and environmental monitoring tasks \cite{zhang2021enabling, wang2024integration}.  A comparison of different ISAC sensing types is given in Table \ref{tab:ISAC_Sensing_Types}.

\vspace{-10pt}
\subsection{RIS-Assisted PLS for ISAC Systems}

The integration of RIS into PLS for ISAC systems offers a novel way to enhance wireless communication security. RIS enables intelligent control of electromagnetic waves, optimizing both sensing and communication functionalities, thus improving security in ISAC systems. In conventional ISAC systems, C\&S tasks are often treated separately. RIS, however, unifies these functions by allowing the communication system to also act as a sensor, utilizing radio signals to monitor the environment. This dynamic adaptation of transmission characteristics helps create secure communication links while supporting sensing tasks like object detection and localization.

RIS enhances PLS by exploiting natural channel imperfections such as noise and fading, which are typically seen as obstacles. By leveraging these imperfections, RIS makes interception by Eves more difficult while maintaining high-quality communication for legitimate users. This ability to control signal propagation ensures that data is securely transmitted, minimizing the risk of interception.
Furthermore, RIS assists in advanced interference management by introducing deliberate noise or jamming signals to disrupt eavesdropping efforts, especially in environments where traditional security methods may fall short. The reciprocity of the wireless channel also allows RIS to generate shared secret keys based on channel characteristics, adding another layer of security.
Additionally, RIS supports the use of diversity techniques—spatial, temporal, and frequency diversity—that enhance the robustness and security of the system. These techniques ensure reliable communication even in challenging environments, such as urban or rural areas, where channel conditions fluctuate.


\subsubsection*{\textbf{Advantages of RIS-Assisted PLS for ISAC Systems}}

RIS-assisted PLS significantly enhances the security and performance of ISAC systems. By exploiting the natural imperfections of wireless channels, such as noise, fading, and interference, RIS improves the secrecy capacity, ensuring that communication remains secure even in challenging environments. RIS allows for real-time control of the phase and amplitude of signals, making it difficult for unauthorized entities to intercept the communication while providing high-quality service to legitimate users. 

Additionally, RIS enables advanced interference management, where controlled noise or jamming signals are strategically introduced to obstruct eavesdropping without affecting the legitimate transmission. The system also benefits from key generation techniques based on the reciprocity of the wireless channel, ensuring secure communication through the generation of shared secret keys. Moreover, RIS employs diversity techniques, including spatial, temporal, and frequency diversity, to further enhance the reliability and robustness of the communication link.

RIS-assisted PLS is scalable and adaptable to dynamic environments, such as urban areas and mobile networks, ensuring robust security even in rapidly changing conditions. Its EE, due to the passive nature of RIS, makes it ideal for low-power devices, while its cost-effectiveness results from simpler hardware compared to traditional systems. Moreover, RIS enables seamless integration of C\&S functions, optimizing resource usage and improving system efficiency. With its capability to dynamically adjust signal propagation, RIS enhances coverage and overall system performance, even in environments where traditional systems struggle.

\vspace{-10pt}
\section{Passive RIS-based Advanced PLS Techniques in ISAC Systems} \label{sec:sectionIII}
In ISAC systems, passive RIS has emerged as a key enabler to improve PLS while optimizing both C\&S capabilities. By intelligently manipulating the propagation of the wireless signal, passive RIS can improve security, increase detection precision, and enhance overall system performance. This section delves into advanced PLS techniques that leverage passive RIS to achieve optimal ISAC operation under various constraints. 

A fundamental challenge in ISAC systems is to maximize system throughput while meeting sensing requirements. Achieving high data transmission rates without compromising sensing accuracy requires careful resource allocation and optimization strategies. At the same time, improving SRs under sensing constraints is crucial for securing wireless communication against Eves. Passive RIS can be strategically deployed to reduce information leakage, ensuring secure transmissions while maintaining effective sensing operations. Another critical aspect is improving sensing SNR while maintaining communication security. Since sensing accuracy is highly dependent on SNR levels, RIS-based techniques can be designed to enhance sensing capabilities without weakening security measures. Furthermore, optimizing the WSR and sensing performance under constraints allows for a balanced trade-off between communication efficiency and sensing precision. By formulating joint optimization models, ISAC systems can dynamically adjust to diverse operational demands.
Moreover, the emergence of STAR-RIS and transmissive RIS-based PLS schemes introduces new dimensions to ISAC security. Unlike traditional RIS, which only reflects signals, STAR-RIS enables more flexible signal control by allowing simultaneous transmission and reflection. This added flexibility can be leveraged to develop more advanced security mechanisms, further strengthening ISAC system resilience. In general, passive RIS-based PLS techniques provide a powerful framework to improve ISAC system security and efficiency. By optimizing system throughput, SRs, sensing SNR, and the trade-off between sensing and communication, while leveraging innovations such as STAR-RIS, ISAC networks can achieve robust, adaptable, and future-ready performance.

\vspace{-8pt}
\subsection{Optimizing System Throughput Under Sensing and Other Constraints} This subsection explores optimization strategies to enhance throughput in RIS-assisted ISAC systems, focusing on secure communication and effective sensing. Key innovations, including Movable Antennas (MAs), RIS-based backscatter systems, UAV-assisted networks, and DISCO-RIS with passive jamming, aim to boost communication rates while mitigating eavesdropping threats. Optimization techniques such as beamforming, RIS reflection coefficient tuning, phase shift adjustments, UAV trajectory planning, and power allocation address complex non-convex challenges. To efficiently solve these challenges, methods like Alternating Optimization (AO), SCA, and manifold optimization are employed. Results demonstrate that adaptive frameworks significantly improve SRs, sensing accuracy, and overall communication efficiency in dynamic ISAC environments.

\begin{figure}
\centering
\includegraphics[width=0.40\textwidth]{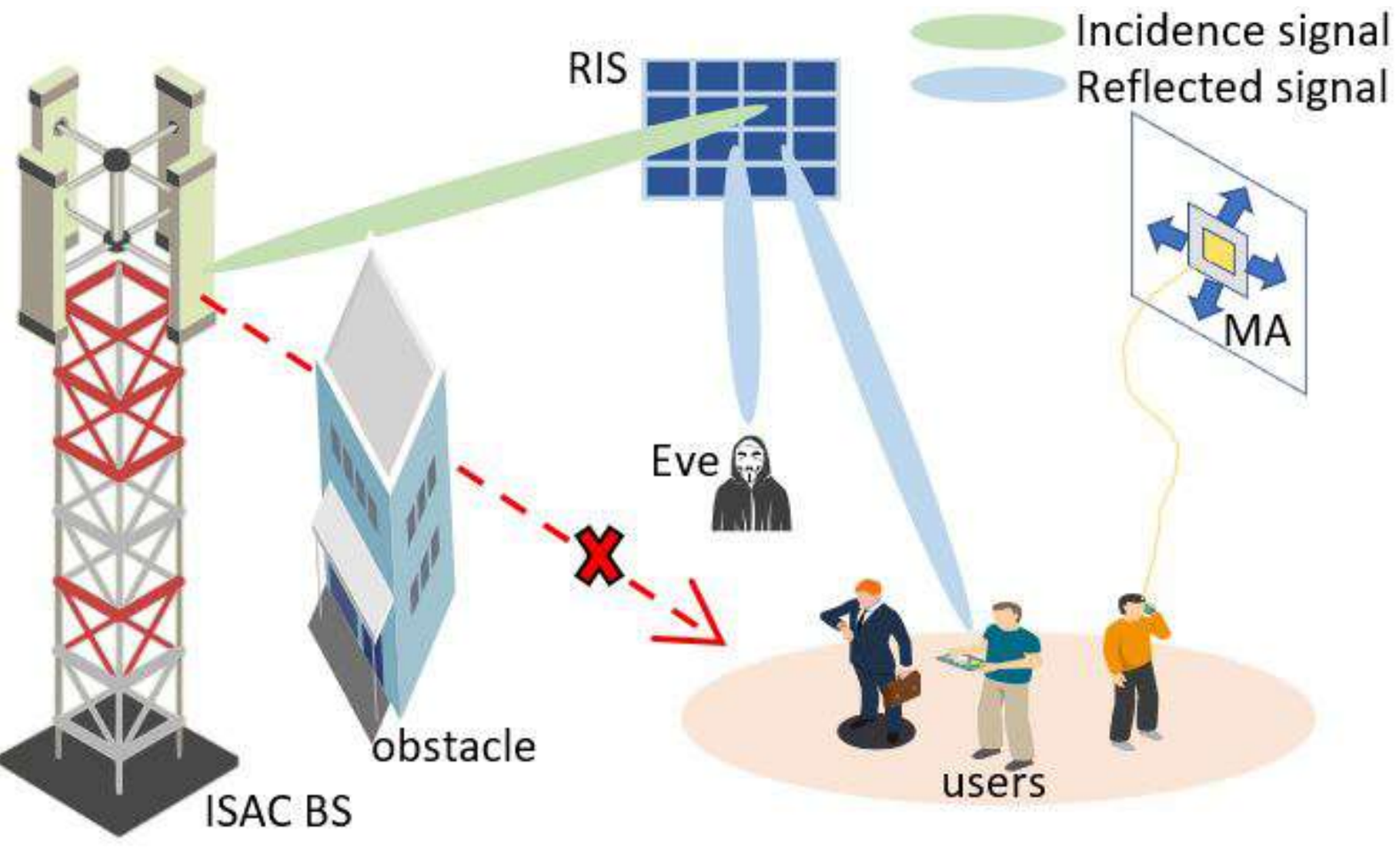}
\caption{Illustration of RIS-ISAC system, where a dual-function BS transmits information to K single-antenna MA users.}
\label{fig:sys throughput}
\end{figure}

A key approach to enhancing ISAC security involves integrating MAs with RIS, as explored in \cite{ma2024movable}. This work proposes an optimization strategy to maximize user sum rates by jointly optimizing BS beamformers, RIS Reflection Coefficients (RCs), and MA locations, as shown in Fig. \ref{fig:sys throughput}. The goal is to balance communication performance, radar sensing efficiency, and eavesdropping defense. To achieve this, a two-layer penalty-based method is introduced, where the outer layer ensures feasibility, and the inner layer applies Lagrange duality for semi-closed-form solutions. Additionally, the Rayleigh quotient method optimizes the BS’s receive beamformer, while a convex subproblem refines the transmit beamformer matrix. Finally, the Majorization-Minimization (MM) algorithm fine-tunes RIS RCs and MA positions. Simulation results confirm that integrating MAs in RIS-ISAC systems significantly enhances security compared to fixed-position antenna systems. Building on this, \cite{xia2024joint} explores RIS-based backscatter systems for secure transmission with sensing support. This study examines a scenario with an Aerial Eve (AE) and multiple IoT devices communicating with an ISAC BS. A novel electromagnetic waveform design is proposed, incorporating prior sensing data to estimate the AE’s angle using the Cramer-Rao lower bound. The objective is to maximize the sum-rate while preventing information leakage through optimized RIS phase shifts. Given the complexity of variable coupling, a Fractional Programming (FP) based AO algorithm is introduced, using SCA and manifold optimization to handle non-convex constraints. Simulation results demonstrate that accurate AE sensing significantly enhances the sum SR, proving the approach’s effectiveness in securing ISAC networks.

Extending the focus to UAV-assisted ISAC networks, \cite{zhang2024robust} investigates secure transmission using RIS-equipped UAVs, which serve as dual-purpose BSs. These UAVs track J targets while simultaneously serving K communication users, requiring efficient coordination between tracking and communication tasks. The study highlights how RIS improves system performance, enabling simultaneous tracking and secure communication. To achieve this, a comprehensive secure transmission scheme is introduced, optimizing RIS phase shifts, user and target scheduling, transmit power allocation, and UAV trajectory and speed. Simulation results confirm that this strategy effectively strengthens security and improves resilience against eavesdropping threats. Adding another dimension to ISAC security, \cite{huang2024integrated} examines the disruptive effects of DISCO jamming attacks, where a DRIS with unpredictable, time-varying reflections accelerates channel aging and disrupts ISAC operations. This work formulates an ISAC problem and proposes a waveform optimization strategy that balances S\&C performance through Pareto optimization. Theoretical analyses quantify the impact of DISCO jamming, while numerical evaluations validate the performance degradation caused by these attacks.
\subsubsection*{Discussion} The reviewed studies demonstrate significant progress in passive RIS-assisted ISAC systems, as shown in Table \ref{T:SYS TH}, focusing on secure communication, sensing accuracy, and throughput optimization. Various techniques, such as beamforming, RIS reflection tuning, UAV trajectory control, and power allocation have been employed. For instance, \cite{ma2024movable} integrates MAs with RIS to improve SRs, while \cite{xia2024joint} enhances RIS-based backscatter systems to mitigate eavesdropping by optimizing waveform design. Similarly, \cite{zhang2024robust} investigates UAV-assisted RIS networks, illustrating their capability to facilitate simultaneous tracking and secure transmission, whereas \cite{huang2024integrated} explores the disruptive effects of DISCO jamming on ISAC performance.
Despite these advancements, several challenges remain unaddressed. Most studies assume static system parameters, overlooking the dynamic nature of ISAC environments, where CSI variations, user mobility, and interference require real-time adaptability. Future research should incorporate learning-based optimization techniques, such as DRL and online learning, to enhance system resilience to environmental fluctuations. Additionally, while \cite{huang2024integrated} addresses jamming threats, a more comprehensive examination of broader adversarial risks, including coordinated eavesdropping and adversarial manipulation of RIS, is warranted. Another limitation is the scalability of current models, as most research focuses on single-RIS, single-UAV, or limited-user scenarios. Expanding inquiry into multi-RIS, multi-user, and multi-UAV networks is essential for enhancing system coordination and operational efficiency. Furthermore, EE and hardware limitations are often overlooked, with many studies assuming ideal RIS phase tuning and unlimited power availability. Future work should adopt hardware-aware optimizations that consider quantized phase shifts, hardware impairments, and low-complexity RIS configurations, bridging the divide between theoretical models and practical applications.
Addressing these gaps is critical for the advancement of next-generation ISAC systems. Future research should concentrate on adaptive learning-driven optimization (\cite{ma2024movable, xia2024joint}), secure ISAC architectures against intelligent threats (\cite{huang2024integrated}), and scalable multi-agent RIS coordination (\cite{zhang2024robust}). Moreover, exploring energy-efficient ISAC frameworks will be vital for balancing power consumption, security, and sensing performance. These enhancements will facilitate the development of resilient, secure, and high-performance ISAC systems, paving the way for 6G and beyond.
\begin{table*}[htbp]
\caption{Summary of System Throughput Under Sensing and Other Constraints Schemes}
\label{T:SYS TH}
\centering
\begin{tabular}{|l|p{0.55cm}|p{1.3cm}|p{.5cm}|p{.5cm}|p{.5cm}|p{1.1cm}|p{1.8cm}|p{1cm}|p{1.8cm}|p{1.5cm}| p{2.5cm}| }
\hline
\multirow{3}{*}{\textbf{Ref.}}   & \multirow{3}{*}{\textbf{Year}} & \multicolumn{2}{c|}{\textbf{System Model}} & \multicolumn{4}{c|}{\textbf{Communication and Sensing Details}} & \multirow{3}{*}{\textbf{Eve's CSI}} & \multirow{3}{*}{\textbf{Methodology}}     & \multirow{3}{*}{\textbf{Opt. Variables}} & \multirow{3}{*}{\textbf{Objective}} \\ \cline{3-8}
& & \textbf{Scenario} & \textbf{\# (RISs)} & \textbf{\# (UEs)} & \textbf{\# (Eves)} & \textbf{Sensing Type} & \textbf{Sensing Metric} & & & & \\ \hline

\cite{ma2024movable} & 2024 & RIS-MAs-ISAC  & S & M &  S &  Eve Detection  & Eve min. sensing SNR  &  Avail. & FP-penalty, Lagrange duality, MM & BS's Tx/Rx BF, RIS's RCs, and UEs' MAs' position  & To maximize users' sum rate. \\ \hline


\cite{xia2024joint} & 2024 &  RIS-BC-ISAC & M & M &  S &   Aerial Eve
sensing & Eve sensing SNR, CRLB & Avail. &  AO, SCA, MO, CRLB & Tx waveform and RIS PS & To maximize sum rate.  \\ \hline
\cite{zhang2024robust} & 2024 & RIS-UAV-ISAC & S & M &  S & Eve detection \& Targets sensing & ASR, AAR & NA. &  AO, SCA, MO & Tx PA, UEs' scheduling, RIS PSs, and UAV trajectory \&
velocity   & To maximize the average achievable rate. \\ \hline
\cite{huang2024integrated} & 2024 & DRIS-FPJ  & S & M & S & Target sensing (UAV) &  Sensing SINR & Perfect & Pareto optimization  &  sensing waveform SINR, DRIS PSs \& amplitudes & To characterize the impact of DISCO
jamming attacks on ISAC systems.  \\ \hline


\multicolumn{12}{l}{\scriptsize {\textit{Ref. - Reference, opt. - optimization , S- Single, M- Multiple, max. - maximization, min. - minimize, BF - beamforming, ASR - average SR, NA. - Not Available }}} \\
\multicolumn{11}{l}{\scriptsize {\textit{ AAR - average achievable rate,  Avail. - available, PSs - phase shifts, RCs - Reflection Coefficients, TX\& RX- Transmit and Receive, PA - power allocation }}} \\
\multicolumn{11}{l}{\scriptsize {\textit{ }}} \\
\end{tabular}
\end{table*}

 \vspace{-10pt}
\subsection{Maximizing Covert and Secrecy Rates Under Sensing and Other Constraints}
This subsection discusses strategies for improving secrecy and covert communication in RIS-assisted ISAC systems without compromising sensing performance. Methods such as beamforming optimization, phase shift control, and trajectory design utilize RIS, AN, and UAV support to mitigate eavesdropping risks. Furthermore, STAR-RIS and TRIS architectures enhance secure communication through dynamic signal management. The subsection is divided into three areas: Covert Rate Under Sensing and Other Constraints Schemes, which address covert transmission with RIS; Reflective-RIS Based SR Under Sensing and Other Constraints Schemes, which focus on secure transmission via RIS-based passive beamforming; and STAR-RIS and TRIS Based SR Under Sensing and Other Constraints Schemes, which emphasize their importance in enhancing PLS. These developments highlight the significance of RIS, AI optimization, and advanced signal processing in fortifying ISAC systems while ensuring efficiency.

\subsubsection{Covert Rate Under Sensing and Other Constraints Schemes}

\begin{figure}
\centering
\includegraphics[width=0.4\textwidth]{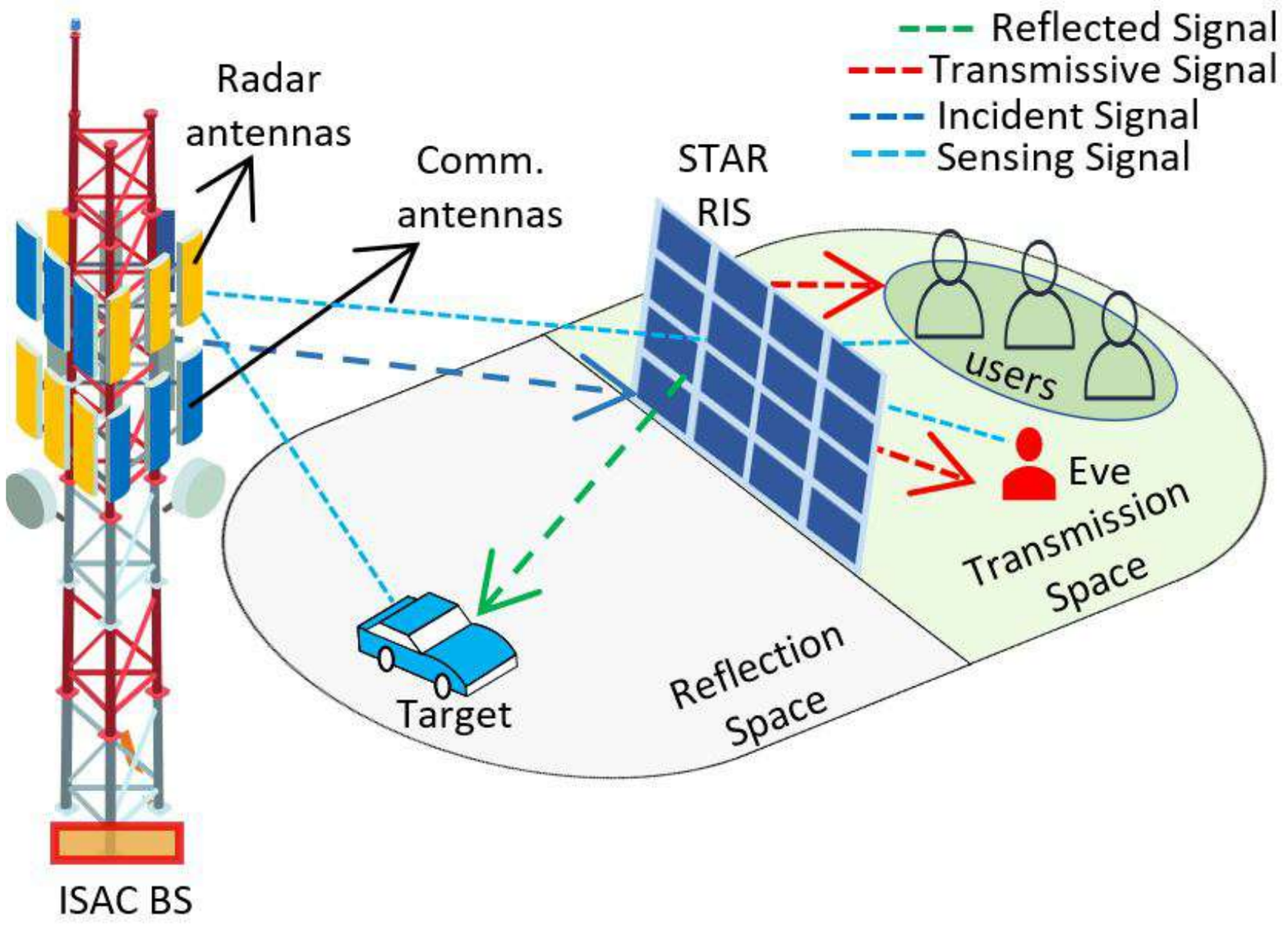}
\caption{STAR-RIS-enabled ISAC network, integrating radar antennas for target sensing and communication antennas for covert communication.}
\label{fig:STAR-RIS}
\end{figure}

In this manuscript \cite{lin2023intelligent}, an ISAC-based secret transmission structure utilizing an RIS and radar for hidden target detection and spontaneous communication is proposed. A combined structure is introduced to improve the secret rate while considering secrecy, total power, and mutual detection constraints. This system includes RIS phase shifts and radar transmit beamforming paths. An SDR technique is employed to find a solution to this challenge. Mathematical results show the effectiveness of the RIS-based secret transmission system, highlighting the trade-off between communication secrecy and radar detection. In this article \cite{liu2024exploiting}, the vulnerabilities of ISAC to wiretapped attacks are assessed. Despite progress, ISAC faces risks due to insufficient security evaluations. The research explores the role of STAR-RIS in secure communication within millimeter-wave ISAC networks, emphasizing imperfect CSI, as shown in  Fig. \ref{fig:STAR-RIS}. A base station (Alice) organizes its multi-antenna into two clusters: one for sensing a target and another for private messaging to a user (Bob) while evading detection from a warden (Willie). STAR-RIS is utilized to create channel uncertainty for Willie, ensuring secure transmission to Bob by adjusting the radio environment. A novel covert transmission method is introduced, where Alice leverages channel uncertainties to obscure Bob's communication. The minimal Probability of Detection Error (PDE) for Willie is analyzed using the Saleh-Valenzuela fading model. The problem is reformulated through S-process and sign-definiteness, resulting in an efficient iterative approach that applies successive convex approximation and penalty techniques. Numerical results validate the theoretical claims, showing significant improvement in covert communication with the proposed solution compared to baseline systems. 

\subsubsection{Refelctive-RIS based Secrecy Rate Under Sensing and Other Constraints Schemes}
This research \cite{chen2023intelligent} discusses robust secure communication in RIS-supported ISAC systems. The setup includes a multi-antenna base station that fulfills radar and communication roles, servicing a single-antenna user while also detecting an Eve. The authors utilize the multiple reflective properties of RIS to enhance secure communication. To address the limited knowledge of instantaneous CSI, the study proposes a robust architecture aimed at maximizing SR. This involves optimizing transmit beamforming and RIS reflection coefficients, factoring in sensing limitations. A bounded uncertainty model is introduced to address challenges related to channel fading inaccuracies and uncertainties regarding Eve's angle. The authors create a mathematical framework for establishing a tractable bound on the combined uncertainty. Additionally, they apply a block coordinate descent technique to tackle the non-convex optimization problem. Simulation results demonstrate that the proposed approach effectively maintains secure communication, even with imperfect Eve CSI.

In this paper \cite{zhao2023joint} the authors investigate ISAC-based RIS-supported PLS. They leverage the radar's powerful signal to distract Eves, aiming to reduce the SINR during unauthorized access. A combined beamforming optimization method is introduced, which integrates RIS passive and  BS active beamforming. The original problem is divided into two subproblems, with the authors using SDP and a FP approach for AO. Simulation results demonstrate improved performance while effectively lowering Eve's SINR. Additionally, the FP-SDP-AO algorithm maintains sufficient quality of service for legitimate users. The authors in \cite{xing2023reconfigurable}  examine information leakage in ISAC caused by eavesdropping during environmental sensing using waveforms. They introduce a RIS-supported beamforming design to enhance PLS within the ISAC framework. The analysis considers an unknown wireless channel with a potential Eve, providing mathematical computations for the approximate Ergodic Attainable Secrecy Rate (EASR). To optimize EASR while ensuring minimum performance for sensing and legitimate communications, a new optimization approach is developed. This method alternates between solutions for the transmit beamformer, AN covariance matrix, RIS phase-shift matrix, and receive beamformer, addressing the problem's non-convex nature. A lower-bound surrogate function is developed to enhance the first two components, employing an SCA method for optimizing the RIS phase-shift matrix. Additionally, a Rayleigh-quotient subproblem is tackled to improve the receive beamformer. Overall, the investigation offers a comprehensive enhancement method with proven theoretical convergence, and through simulation results, it demonstrates that the proposed framework surpasses existing benchmarks in both security and sensing performance, confirming the estimated EASR. 
Another research effort \cite{jiang2023secure} investigates ISAC and eavesdropping threats related to communication messages from specific targets, highlighting wireless security challenges. It assesses PLS in ISAC and the eavesdropping threats linked to communication messages from specific targets, bringing to light various wireless security challenges. It assesses PLS in ISAC systems with the utilization of RIS technology. The authors address the non-convex optimization problem through alternating optimization and successive convex approximation, effectively breaking it down into two convex sub-problems. To facilitate the solution, both SDR and Taylor expansion techniques are employed. Numerical simulations further validate the effectiveness of their proposed algorithm in enhancing security. In a different investigation \cite{liu2023drl}, a RIS-based ISAC system is formulated to improve PLS while simultaneously monitoring targets and serving multiple users. In this context, AN is strategically employed as a defense mechanism against Eves, who are treated as radar targets attempting to capture user information. The main objective is to develop a unified design that integrates RIS phase-shift settings, AN signals, and beamforming strategies for legitimate users, ultimately enhancing the achievable SR. The results from the DRL algorithm show notable enhancements in the SR compared to benchmark methods.

\begin{figure}
\centering
\includegraphics[width=0.40\textwidth]{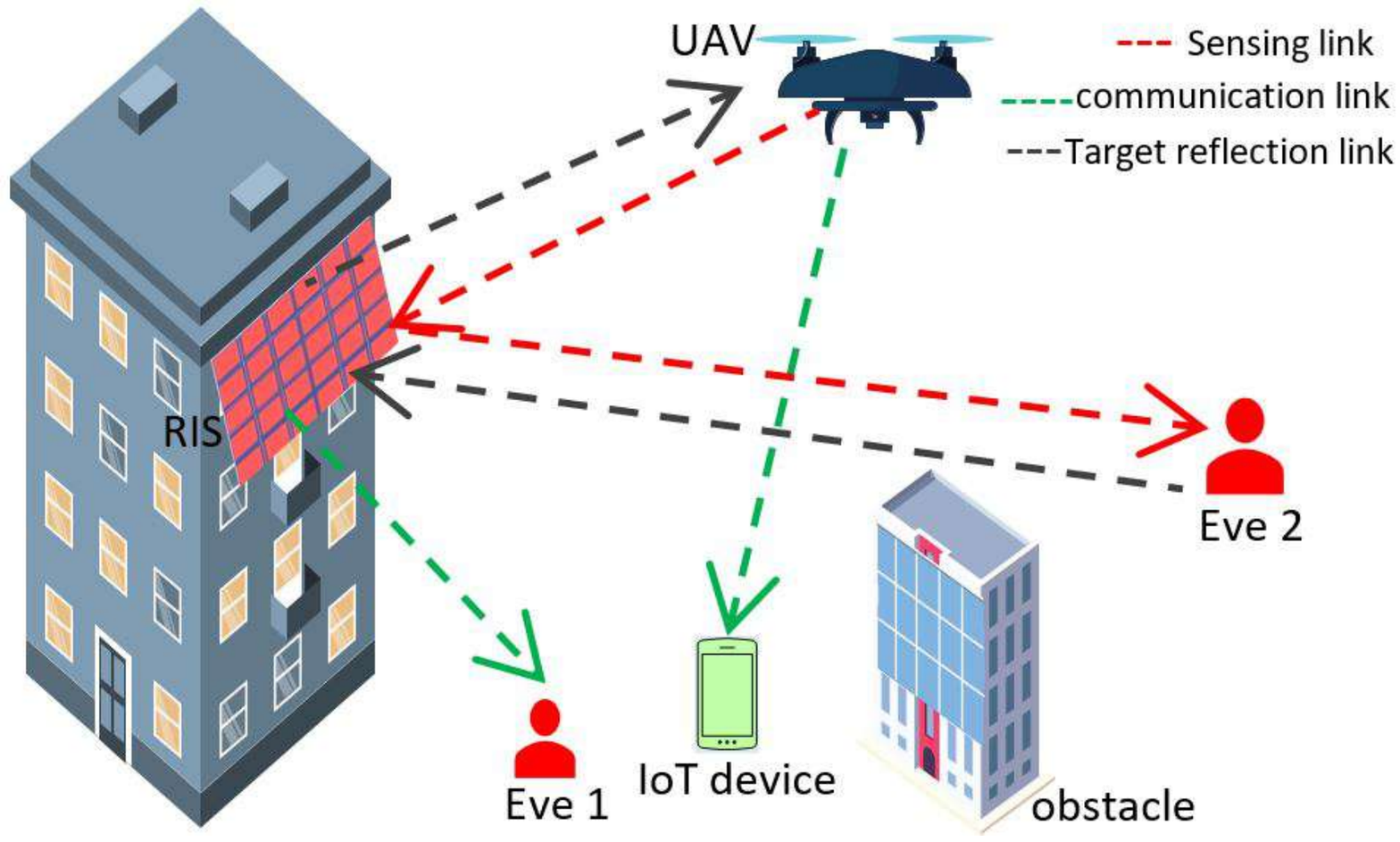}
\caption{RIS-assisted ISAC-UAV system transmitting communications to an IoT device across multiple eavesdroppers.}
\label{fig:maximizing secrecy rate}
\end{figure}

In this work \cite{jiang2024ris}, the ISAC system emphasizes the role of RIS in enhancing security. RIS is vital for secure user data transmission within the ISAC framework. The objective is to improve user secure rates by optimizing target sensing through beamforming and RIS's discrete phase shifter. However, linking optimization variables to conventional methods is difficult. To address this, beamforming and RIS phase shifters are integrated into a DRL framework using Soft Actor-Critic (SAC) and AO algorithms. Simulation results demonstrate that the proposed approach significantly improves user SRs compared to benchmark techniques. In this study \cite{wang2024secure}, the authors introduce a dual-functional radar and communication base station that serves a multi-antenna communication user while detecting hostile eavesdropping targets. They first establish the system model, detailing the signal models for both radar target identification and communication. To tackle the non-convex constraint issue, an approximation technique is utilized, transforming it into an alternating optimization problem. Auxiliary variables are introduced, allowing for the optimization of each block's variables through a block coordinate descent method. The effectiveness of this approach is validated by simulation results, which indicate that the proposed method significantly enhances the system's SR while also achieving excellent target identification outcomes.
In  \cite{xiu2024secure}, the authors explore an ISAC system enhanced by RIS to strengthen security performance in UAV communications. They employ a multiple-antenna UAV to transmit ISAC waveforms while detecting suspicious targets in the presence of an Eve interacting with IoT devices, as shown in  Fig. \ref{fig:maximizing secrecy rate}. To tackle this challenge, a BCD method is proposed for optimization. Additionally, to address four distinct irregular subproblems, an SDR approach, coupled with an SCA algorithm, is implemented. The mathematical results underscore the effectiveness of this approach, indicating that the algorithm not only improves target sensing accuracy but also enhances SRs for IoT devices, thereby reinforcing the overall security of the communication system. 

In this work \cite{jiang2025ris}, resilient PLS techniques for RIS-assisted ISAC systems are analyzed. PLS design utilizes sensing data from ISAC, laying a crucial foundation for enhancing communication and sensing efficiency. To optimize the sum SR while adhering to user QoS constraints, power limits, and sensing requirements, a RIS-assisted joint active and passive beamforming design problem is formulated. This innovative approach considers imperfect sensing estimation and employs radar signals as AN, integrating security with operational efficiency. Initial security approximations establish constraints for Eve's CSI as imperfect, serving as a basis for further analysis. To refine the model, symbolic-deterministic methods and the S-procedure are employed to adjust the infinite inequalities, which ensures robustness in the design. Simulation results indicate that the sensing functionality significantly enhances security and effectively demonstrates the proposed system's capability in balancing C\&S quality.

 \begin{table*}[htbp]
\caption{Summary of Maximizing Covert and Secrecy Rates Under Sensing and Other Constraints Schemes}
\label{T:Covert}
\centering
\begin{tabular}{|p{0.55cm}|l|p{0.55cm}|p{1.3cm}|p{.5cm}|p{.5cm}|p{.5cm}|p{1.1cm}|p{1.2cm}|p{1cm}|p{1.8cm}|p{1.5cm}| p{2.5cm}| }
\hline
\multirow{3}{*}{\textbf{{Cat.}}} & \multirow{3}{*}{\textbf{Ref.}}   & \multirow{3}{*}{\textbf{Year}} & \multicolumn{2}{c|}{\textbf{System Model}} & \multicolumn{4}{c|}{\textbf{Communication and Sensing Details}} & \multirow{3}{*}{\textbf{Eve's CSI}} & \multirow{3}{*}{\textbf{Methodology}}     & \multirow{3}{*}{\textbf{Opt. Variables}} & \multirow{3}{*}{\textbf{Objective}} \\ \cline{4-9}
& & & \textbf{Scenario} & \textbf{\# (RISs)} & \textbf{\# (UEs)} & \textbf{\# (Eves)} & \textbf{Sensing Type} & \textbf{Sensing Metric} & & & & \\ \hline

\multirow{2}{*}{\rotatebox{90}{\shortstack{Covert Rate \\ Schemes}}} & \cite{lin2023intelligent} & 2023 & RIS-ISAC (Covert) & S  & S &  S &  Target detection & Detection mutual info. & Imperfect & AO, SDR (SLBM) &  Tx BF, RIS PSs  & To maximize covert rate.  \\ \cline{2-13}
& \cite{liu2024exploiting} & 2024 &  STAR-RIS-ISAC & S & S &  S &  Target sensing & Minimum  PDE  & Imperfect & S-procedure, SCA, PCCP  & Tx BF and STAR-RIS Tx RCs  & To maximize covert rate while ensuring sensing accuracy.  \\ \hline\hline

 & \cite{chen2023intelligent}  & 2023 & RIS-ISAC & S & S &  S & Eve sensing & Sensing power or accuracy & Imperfect &  Bounded uncertainty model, BCD  & Tx BF, RIS RCs  &  To maximize the SR. \\ \cline{2-13}

& \cite{zhao2023joint} & 2023 & RIS-ISAC & S & M &  S & Target and Eve sensing & Eve SINR  & Avail. & FP-SDP-AO & Active BF,  RIS PS & To minimize the maximum eavesdropping SINR while ensuring QoS for users.  \\ \cline{2-13}

& \cite{xing2023reconfigurable} & 2023 & RIS-ISAC  & S & S &  S & Target and  Eve Detection &  Target min. SNR & NA. & SCA, Rayleigh-Quotient Opt., Surrogate lower-bound construction & Tx BF, AN, RIS PSs, Rx BF   & To maximize EASR with minimum C\&S requirements.  \\ \cline{2-13}
\multirow{9}{*}{\rotatebox{90}{Reflective-RIS based SR Schemes}}
& \cite{jiang2023secure} & 2023 & RIS-ISAC  & S & M &  S & Target sensing &  Target min. SINR & Avail. & AO, SCA,  Taylor expansion and SDR & Tx BF, AN, RIS PSs & To maximize multi-user sum SR. \\ \cline{2-13}

& \cite{liu2023drl} & 2023 & RIS-ISAC & S & M & S & Target detection & Target SINR & Imperfect & SAC based DRL & Tx BF, AN , RIS PSs & To maximize achievable SR.  \\ \cline{2-13}

& \cite{jiang2024ris} & 2024 & RIS-ISAC  & S & S &  M & Targets sensing & Targets sensing perf. & Avail. & DRL, SAC, AO  & Tx BF, RIS Discrete PSs & To maximize secure rates, ensure sensing performance.  \\ \cline{2-13}

& \cite{wang2024secure} & 2024 & RIS-ISAC & S & S &  M & Multi-target sensing & Target detection &  Avail. & AO, BCD & Tx BF, RIS PS & To maximize the SR, while ensuring high target detection performance. \\ \cline{2-13}

& \cite{xiu2024secure} & 2024 & RIS-UAV-ISAC & S & S &  M & Target   detection & Target SNR & Imperfect & BCD, SCA, SDR, MM & Tx BF, RIS PSs, UAV trajectory and Rx BF & To maximize the average communication SR. \\ \cline{2-13}

& \cite{jiang2025ris} & 2025 &  RIS-assisted ISAC system  & S & M & S & Target sensing  & Beampattern gain & Imperfect &  SOC , SCA, AO & Active BF, radar Signal BF, RIS PSs & To maximize sum SR.  \\ \hline\hline

 & \cite{liu2023exploiting} & 2023 & STAR-RIS-ISAC & S & M &  S & Target detection & Sensing SNR & NA. & SCA, SRCR & BS Tx BF, STAR-RIS Tx \& Rx  BF  & To maximize SR, ensure sensing SNR. \\ \cline{2-13}
\multirow{5}{*}{\rotatebox{90}{STAR-RIS and TRIS based SR Schemes}}
& \cite{zhu2023drl} & 2023 & STAR-RIS-ISAC & S & S & S & Target sensing & min. SNR  & Perfect & DRL (DDPG and (SAC)  & BS BF, Rx filters, STAR-RIS Tx and RCs & To maximize SR while ensuring sensing SNR and communication constraints.  \\ \cline{2-13}

& \cite{zhu2024ai} & 2024 &  STAR-RIS-ISAC  & S & M &  M &  Target detection  & Echo SNR  & Perfect & DDPG \&  SAC & BS's Tx BF, Rx filters, STAR-RIS Tx RCs & To maximize the sum SR for LU while ensuring echo SNR thresholds and LU rate constraints. \\ \cline{2-13}

& \cite{wei2024joint} & 2024 & STAR-RIS-ISAC & S & M &  S &  Target sensing & min. beampattern gain  & Avail. & SCA, AO & Tx BF, AN, STAR-RIS Tx \& reflection BF  & To maximize SR and ensure target sensing accuracy.  \\ \cline{2-13}

& \cite{liu2024enhancing} & 2024 &  TRIS-ISAC & S & M &  M &  Eve detection & Secrecy Outage Probability, CRB  & Imperfect & RSMA, BCD, SOCP, S-Procedure, Bernstein’s inequality, SCA   & Common \& Private stream BF, Timeslot Duration  & To maximize SR, improve SEE, minimize CRB. \\ \hline
\end{tabular}
\end{table*}

\subsubsection{STAR-RIS and TRIS based Secrecy Rate Under Sensing and Other Constraints Schemes}
This subsection covers enhancements in ISAC systems for PLS under sensing constraints using STAR-RIS and TRIS frameworks. STAR-RIS improves multi-user C\&S accuracy through optimized beamforming. Covert transmission techniques leverage channel uncertainties to hide signals from Eves, using methods like successive convex approximation. STAR-RIS-assisted ISAC systems increase SRs through artificial jamming and optimized passive reflection, applying alternating optimization strategies. DRL-based methods adjust base station beamforming and STAR-RIS phase shifts dynamically, offering improved security over traditional RIS methods. TRIS-based ISAC transceivers utilize time-division sensing-communication and Rate Splitting Multiple Access (RSMA) to mitigate eavesdropping risks, optimizing beamforming with second-order cone programming and successive convex approximation. STAR-RIS further enhances security by optimizing phase-shift matrices to prevent data leakage while improving sensing accuracy. These advancements underscore the importance of STAR-RIS, TRIS, and various optimization techniques in enhancing the security and efficiency of ISAC systems.

In this study \cite{liu2023exploiting}, the authors enhance PLS by integrating a STAR-RIS. The BS transmits a communication signal along with a sensing signal for single target detection and multi-user communication. The STAR-RIS improves signal propagation, communication, and sensing quality while reducing eavesdropping risks. The main aim is to optimize the overall SR by refining BS transmit beamforming with STAR-RIS transmission and reflection beamforming, while meeting sensing SNR requirements. In order to tackle the non-convex optimization challenges associated with this optimization, the authors employ sequential rank-one constraint relaxation and SCA techniques. This approach enhances network adaptability, ensuring full-space coverage. 
In \cite{zhu2023drl}, the authors examine secure communication in an ISAC system utilizing a STAR-RIS. They enhance the Legitimate User's (LU) average long-term security rate by optimizing the BS transmit beamforming, receive filters, and STAR-RIS coefficients for both transmission and reflection. This enhancement maintains the SNR of the echo's lower bound and complies with the LU's achievable rate limits. To address the non-convex optimization challenge for long-term benefits, the authors propose two DRL algorithms for optimizing BS beamforming and STAR-RIS phase shifts. Simulation results show that STAR-RIS outperforms traditional RIS based on two benchmarks, and provide a detailed assessment of the DRL algorithms' effectiveness.

In  \cite{zhu2024ai}, STAR-RIS uses time-switching protocols and energy splitting, aligning LUs and targets oppositely. Combining BS transmit beamforming with STAR-RIS's coefficients improves the LUs' long-term average security rate. This approach effectively addresses echo signal-to-noise ratio parameters and LUs' rate limits. Traditional convex optimization struggles under changing channel conditions, resulting in high computational complexity and reduced system performance. To manage this complex non-convex challenge and continuity control, reinforcement learning techniques like SAC and Deep Deterministic Policy Gradient (DDPG) are implemented. Simulation results reveal that STAR-RIS consistently outperforms benchmarks in RIS and ISAC systems. 
In this research \cite{wei2024joint}, the authors tackle secure transmission in a STAR-RIS aided ISAC system.  The optimization involves transmit beamforming, AN jamming, and STAR-RIS beamforming to enhance SRs while reducing beam pattern gain for the target. To tackle the non-convex problem, the authors decompose it into two subproblems and employ successive convex approximation. In another work \cite{liu2024enhancing}, This paper proposes a TRIS-enabled transceiver to enhance the security and efficiency of ISAC networks. A time-division  mechanism is designed for resource sharing between sensing and communication. To mitigate interference and prevent eavesdropping, RSMA is utilized, where the common stream  serves as both a signal and AN. The model accounts for imperfect CSI and provides a tight error bound for unauthorized users.
A secrecy sum-rate optimization problem is formulated to enhance security by optimizing beamforming ( matrices for both CS and private stream, along with timeslot  allocation. Due to the problem's non-convexity, a BCD-based second-order cone programming (SOCP) approach is applied, solving subproblems iteratively until convergence. Methods such as the S-procedure, Bernstein’s inequality, and SCA are employed to handle constraints. Simulations validate the superiority of the approach in improving Secrecy Energy Efficiency (SEE) and CRB.

\subsubsection*{Discussion} This passive RIS-based subsection, as outlined in Table \ref{T:Covert}, encompasses three subsubsections covering covert rate optimization, reflective-RIS-based secure communication, and STAR-RIS/TRIS-based SR enhancements, utilizing beamforming, RIS phase tuning, UAV trajectory control, power allocation, and AN injection.

Research efforts \cite{lin2023intelligent, liu2024exploiting} concentrate on covert rate maximization in RIS-assisted ISAC. \cite{lin2023intelligent} establishes a RIS-based covert transmission framework, optimizing beamforming and phase shifts while addressing mutual information leakage constraints, under the assumption of imperfect CSI. \cite{liu2024exploiting} utilizes STAR-RIS to generate channel uncertainty and implements SCA and S-procedure-based optimization, while also taking into account imperfect CSI. Nevertheless, both methodologies depend on static models, which are inadequate in dynamic adversarial environments. Future research should incorporate reinforcement learning-based covert transmission for real-time security adaptation.

This reflective RIS-based SR optimization category includes \cite{chen2023intelligent, zhao2023joint, xing2023reconfigurable, jiang2023secure, liu2023drl, jiang2024ris, wang2024secure, xiu2024secure, jiang2025ris}, all optimizing beamforming and RIS RCs to counteract eavesdropping threats. \cite{chen2023intelligent} maximizes SRs using bounded uncertainty models and block coordinate descent, assuming imperfect CSI. \cite{zhao2023joint} applies RIS-assisted PLS, disrupting Eves with radar signals, using FP-SDP-AO under available CSI. \cite{xing2023reconfigurable} proposes an EASR-based secrecy model, integrating SCA and Rayleigh-quotient optimization, assuming no CSI knowledge.
\cite{jiang2023secure} enhances multi-user SRs via SCA, SDR, and Taylor expansion, assuming available CSI, while \cite{liu2023drl} utilizes DRL-based phase-shift learning under imperfect CSI. \cite{jiang2024ris} employs SAC-based DRL to optimize RIS discrete phase shifts, assuming available CSI. \cite{wang2024secure} introduces a multi-target sensing RIS-ISAC model, optimizing AO and BCD for SRs under available CSI. \cite{xiu2024secure} examines RIS-UAV-ISAC security, optimizing UAV trajectories and beamforming via BCD-SCA-SDR-MM, assuming imperfect CSI. \cite{jiang2025ris} maximizes sum SRs in RIS-ISAC, using SCA, AO, and SOC, assuming imperfect CSI. Future research should refine probabilistic CSI models and adversarial ML techniques to enhance resilience against Eve uncertainty.

In STAR-RIS and TRIS-based SR optimization category, studies \cite{liu2023exploiting, zhu2023drl, zhu2024ai, wei2024joint, liu2024enhancing} optimize STAR-RIS and TRIS-based SR performance through beamforming and reflection control. \cite{liu2023exploiting} explores STAR-RIS-aided SR maximization, using SCA-SRCR techniques, with no available Eve CSI. \cite{zhu2023drl} applies DRL-based optimization with DDPG and SAC to improve secrecy while ensuring sensing SNR, assuming perfect CSI. \cite{zhu2024ai} integrates time-switching and energy-splitting for STAR-RIS, leveraging DDPG and SAC-based learning, assuming perfect CSI for long-term secrecy optimization.
\cite{wei2024joint} introduces a joint STAR-RIS optimization model, incorporating AN jamming and beamforming-based SCA-AO, assuming available CSI. \cite{liu2024enhancing} proposes a TRIS-enabled ISAC transceiver, employing RSMA, BCD-SOCP, S-Procedure, and Bernstein’s inequality to enhance secrecy EE, assuming imperfect CSI. Future work should refine adaptive STAR-RIS and TRIS security frameworks by combining probabilistic CSI learning and reinforcement-based optimization.

Key challenges persist, particularly the assumption of available or perfect CSI for Eves, as seen in \cite{zhao2023joint, jiang2023secure, jiang2024ris, wang2024secure, wei2024joint, zhu2023drl, zhu2024ai}. In reality, attackers often operate under unavailable or imperfect CSI, making current security models vulnerable in practical scenarios. Future research should integrate probabilistic CSI models, robust estimation techniques, and adversarial ML frameworks to enhance system robustness against unknown Eves.
Additionally, reliance on offline AO, SDR, and convex optimization limits real-time adaptability, while scalability in multi-agent ISAC networks remains unaddressed (\cite{liu2023drl, zhu2024ai, jiang2024ris, xiu2024secure}). Expanding MARL and decentralized beamforming strategies can improve large-scale ISAC security coordination by dynamically adjusting network parameters in adversarial conditions.
Furthermore, hardware constraints, such as quantized RIS phase shifts, reflection impairments, and power limitations, are often ignored (\cite{liu2023exploiting, wei2024joint, jiang2025ris, liu2024enhancing}). Most studies assume ideal RIS control, overlooking practical deployment challenges. Future work should develop low-complexity, hardware-aware optimization frameworks to ensure feasibility in real-world ISAC implementations.
Lastly, SR maximization often neglects sensing accuracy, necessitating a unified secrecy-sensing trade-off framework (\cite{liu2024enhancing, wang2024secure, xiu2024secure}) to balance secure transmission and target detection. Addressing these gaps will be crucial in enabling scalable, adaptive, and energy-efficient ISAC security solutions for 6G and beyond.

\vspace{-10pt}
\subsection{Maximizing Sensing SNR Under Secrecy and Other Constraints}
This subsection discusses strategies for enhancing sensing SNR in RIS-assisted ISAC systems, emphasizing secure communication. Techniques like joint beamforming, radar filtering, and phase shift optimization enhance sensing accuracy and reduce information leakage. Optimization methods such as SCA, SOCP, and MM refine security in sensing operations. Moreover, virtual LoS links through RIS facilitate secure target detection and communication. These advances underscore the significance of RIS and sophisticated optimization for optimizing sensing SNR while maintaining ISAC security and efficiency.

This work \cite{chu2023joint} investigates an ISAC system improved by RIS. The BS detects malicious radar targets while maintaining secure communication for MU-MISO.
RIS enhances secure wireless communication by transmitting a specific sensing signal alongside the communication signal, which mitigates information leakage and enhances sensing performance. The primary objective is to increase the radar's SNR through optimized transmit and reflection beamforming as well as radar receive filtering, while adhering to power and RIS reflecting coefficient constraints. A novel method is proposed to tackle the non-convex multivariate coupling issue by decomposing it into three independent problems. To address these problems iteratively, the authors employ FP, SDR, and MM algorithms. Simulation results demonstrate that the secure RIS-ISAC system achieves a 2 dB enhancement in radar performance compared to configurations without RIS. The findings demonstrate that the proposed secure RIS-ISAC system offers a notable 2 dB improvement in radar efficacy over scenarios lacking RIS, underscoring its effectiveness and potential in enhancing secure communications and radar target detection. 

The authors in \cite{kumar2023sca} examine a secure RIS-based ISAC system, focusing on optimizing transmission and reflection beamforming patterns to enhance beampattern gain. The non-convex nature of the problem complicates the interrelations among design variables. Addressing these challenges, along with the integration of AO methods in practical applications, remains an ongoing pursuit. An innovative SCA-based SOCP framework is proposed, facilitating the simultaneous update of all design variables. This SCA approach significantly surpasses a previously established penalty-based benchmark method. Moreover, it demonstrates that, although it requires slightly more computational effort, the average problem resolution time is considerably reduced. In a related study \cite{hua2023secure1}, the authors shift focus to RIS-based downlink communication for multiple users, creating a virtual LoS link for target sensing. To combat eavesdropping, the BS transmits specialized sensing signals that sustain the sensing quality. Two scenarios are examined—target location and the presence of perfect CSI—necessitating distinct optimization strategies. In the ideal case where the BS is aware of the target’s location, a penalty-based technique is employed. The MM approach yields closed-form solutions for RIS phase shifts, while Lagrange duality formulates semi-closed-form solutions for beamformers. For imperfect CSI scenarios, a robust method leveraging the S-procedure and sign-definiteness is proposed. Simulation results underscore the effectiveness of the proposed methods in balancing C\&S quality, demonstrating RIS's potential in enhancing sensing capabilities and securing ISAC systems. Similarly, in \cite{sun2023security}, the authors introduce a security-enhanced ISAC system utilizing phase-coupled Intelligent Omni-Surfaces (IOS) to enable simultaneous C\&S without additional sensors. The IOS divides the environment into separate C\&S regions, providing services to multiple users while establishing a virtual LoS link for target detection. Additionally, it enhances PLS by mitigating the risk of information interception by sensing targets.
To limit information leakage, a joint optimization approach is proposed, integrating the design of communication beamformers, sensing beamformer, and IOS phase-shift matrices. The objective is to maximize sensing beam gain while ensuring the required SINR for communication and minimizing leakage to potential Eves. Two alternative optimization algorithms are developed: one based on an independent phase-control model, and the other utilizing a coupled-phase model, both aimed at refining the IOS phase-shift matrices for improved security and sensing efficiency.

\begin{table*}[htbp]
\caption{Summary of Maximizing Sensing SNR Under Secrecy and Other Constraints Schemes}
\label{T:SNR}
\centering
\begin{tabular}{|l|p{0.55cm}|p{1.3cm}|p{.5cm}|p{.5cm}|p{.5cm}|p{1.1cm}|p{1.6cm}|p{1cm}|p{1.8cm}|p{2.3cm}| p{2.5cm}| }
\hline
\multirow{3}{*}{\textbf{Ref.}}   & \multirow{3}{*}{\textbf{Year}} & \multicolumn{2}{c|}{\textbf{System Model}} & \multicolumn{4}{c|}{\textbf{Communication and Sensing Details}} & \multirow{3}{*}{\textbf{Eve's CSI}} & \multirow{3}{*}{\textbf{Methodology}}     & \multirow{3}{*}{\textbf{Opt. Variables}} & \multirow{3}{*}{\textbf{Objective}} \\ \cline{3-8}
& & \textbf{Scenario} & \textbf{\# (RISs)} & \textbf{\# (UEs)} & \textbf{\# (Eves)} & \textbf{Sensing Type} & \textbf{Sensing Metric} & & & & \\ \hline
\cite{chu2023joint} & 2023 & RIS-ISAC & S & M &  S &  Target detection & Target SNR  &  Perfect & AO, SDR, FP, MM &  Tx \& Reflection BF , Radar Rx Filter and RIS RCs & To maximize radar SNR and ensure secure communication. \\ \hline
\cite{kumar2023sca} & 2023 & RIS-ISAC  & S & M &  S &  Target sensing  & Beampattern gain toward target  &  Avail. & SCA, SOCP &  Tx BF, RIS PSs  &  To maximize the beampattern gain at the target. \\ \hline


\cite{hua2023secure1} & 2023 &  RIS-ISAC & S & M &  S & Target sensing & Beampattern gain and min. users SINR & Perfect & Lanrange duality, MM, S-procedure & Comm. \& radar BFs, RIS PSs & To maximize sensing beampattern gain while ensuring security.  \\ \hline

\cite{sun2023security} & 2023 &  IOS-ISAC & S &  M & S & Target
sensing & Sensing Beam Pattern Gain, min. SINR & Avail. & Independent Phase-and Coupled-Phase Models  & Comm. \& Sensing BFs and RIS PS  & To maximize sensing gain, ensure secure communication and SINR.  \\ \hline

 \cite{ye2025joint} & 2025 & RIS-ISAC  & S & M &  M & Targets detection  & MWBG & Avail. & AO, SDR, AN & Tx  BF, RIS PSs, AN design & To maximize minimum sensing performance under SR and power constraints. 
 \\ \hline

 \cite{xu2025intelligent} & 2025 & IS-Radar stealth against unauthorized ISAC & S & S &  M & AOA Sensing & AOA Estimation Error & Avail. & Game-theoretic optimization, closed-form solution & PS array, Reflection channel variable & To maximize AoA distortion. 
 \\ \hline

\end{tabular}
\end{table*}



In \cite{ye2025joint} an investigation explores an RIS-aided ISAC system using PLS, centered on users and targets in NLoS situations relative to the BS. The RIS is crucial for establishing virtual LoS links.  An AO framework is introduced to address this challenge, including a complexity evaluation. Numerical results demonstrate that this approach outperforms random phase and separate beamforming methods in terms of sensing performance. In \cite{xu2025intelligent}, the authors discuss the integration of radar sensors with communication networks in 6G wireless systems, highlighting the accompanying security risks, particularly regarding unauthorized access to user location data via dual-functional base stations (DFBS). To mitigate these concerns, an intelligent surface (IS)-assisted radar stealth technology is proposed. This innovative technology safeguards users by adjusting IS phase shifts to modify wireless channels, thereby ensuring uninterrupted communication. The aim is to amplify the discrepancy between DFBS angle of arrival (AoA) estimates and their actual values, while maintaining minimum SNR for communication. The scenario is conceptualized as a game, where the DFBS attempts to maximize its utility while the IS aims to minimize it. The study demonstrates that the nonconvex optimization problem can be effectively resolved through geometric analysis, resulting in a closed-form solution. Simulations further illustrate that this method exceeds baseline techniques for unauthorized sensing detection while concurrently reducing interference in wireless communication. 

\subsubsection*{Discussion} The Table \ref{T:SNR} summarizes schemes for sensing SNR maximization in RIS-assisted ISAC systems, focusing on secure communication and target sensing through joint beamforming, radar filtering, and RIS phase tuning. Optimization methods like SCA, SOCP, and MM enhance security, while virtual LoS links improve target detection. \cite{chu2023joint, kumar2023sca, hua2023secure1, sun2023security} refine beamforming and RIS RCs to boost sensing accuracy and limit leakage. \cite{chu2023joint} employs AO, SDR, FP, and MM for transmit and reflection beamforming, while \cite{kumar2023sca} applies SCA-SOCP to maximize beampattern gain. \cite{hua2023secure1} integrates penalty-based optimization, MM, and Lagrange duality, and \cite{sun2023security} leverages IOS phase models. Advanced works like \cite{ye2025joint, xu2025intelligent} optimize multi-target sensing, with \cite{ye2025joint} using AO for secrecy-constrained sensing and \cite{xu2025intelligent} applying game-theoretic radar stealth against unauthorized ISAC access.

Despite progress, key challenges persist. Many studies \cite{chu2023joint, hua2023secure1} assume perfect CSI, which is unrealistic; future research should adopt probabilistic CSI models and robust estimation. Additionally, methods like \cite{kumar2023sca, sun2023security, ye2025joint} rely on offline AO, SDR, and SOCP, limiting real-time adaptability; integrating reinforcement learning-based optimization can improve dynamic security responses. Hardware constraints, including RIS phase quantization and power limitations, are often overlooked \cite{kumar2023sca, ye2025joint, xu2025intelligent}, necessitating hardware-aware optimization for practical deployment. Furthermore, most studies focus on either sensing gain or secrecy, but a unified secrecy-sensing trade-off framework is needed for efficient balance. Addressing these gaps will enhance scalability, adaptability, and EE, ensuring robust RIS-assisted ISAC security for next-generation networks.

\vspace{-8pt}
\subsection{Optimizing WSR and Sensing Under Constraints} 
This subsection reviews advanced optimization methods aimed at enhancing WSR and sensing performance in RIS-assisted ISAC systems with a focus on security. Innovations such as DFRC, semi-passive RIS, and secure beamforming are introduced to enhance sensing accuracy and communication security. Techniques like frequency-shifted chirp spread spectrum modulation, symbol-level precoding, and active-passive beamforming optimization are used to counter eavesdropping risks while ensuring efficient communication. Additionally, frameworks including FP, semi-definite programming, and successive convex approximation are utilized to optimize transmission and sensing strategies, effectively handling interference and security concerns. These advancements highlight the crucial impact of RIS, adaptive optimization, and interference-based security measures in enhancing ISAC system performance, ensuring both secrecy and efficient sensing.

This work \cite{jin2023ris} explores the DFRC system enhanced by RIS support, introducing a novel ISAC strategy tailored for future 6G networks. The authors introduce a DFRC design utilizing Frequency-Shifted Chirp Spread Spectrum Index Modulation (FSCSS-IM), facilitating substantial data transfer through radar signals. Additionally, the system employs radar-acquired azimuth information for RIS-assisted beamforming in communication, addressing amplitude and phase distortions from potential Eves. Simulation results reveal that RDFI achieves strong physical layer communication security and enhances Bit Error Rate (BER) performance, reinforcing the effectiveness of RDFI in both aspects. 

In \cite{wei2024cramer}, the authors focus on ISAC as a solution for energy and spectrum shortages by combining C\&S. However, a potential Eve may attempt to intercept the information sent to the communication user. The authors analyze two target types: point and extended targets, where direction of arrival estimation is crucial for point targets, while the complete target response matrix is necessary for extended ones. To enhance efficiency, a weighted optimization problem is formulated to minimize CRB and maximize SRs through a joint optimization of RIS phase shifts and transmit beamforming. Addressing the non-convex problems associated with both target types involves strategies to combat energy and spectrum shortages, employing AO and SCA approximation techniques. Simulation results indicate that the proposed methods outperform existing benchmarks. Similarly, in  \cite{song2024secure}, the authors examine a secure RIS-assisted ISAC system, in which a BS transmits a dual-function waveform to support users while detecting Eves. A symbol-level precoding approach is employed to enhance security and communication performance. The optimization focuses on maximizing user SNR and CRB while adhering to power constraints, constructive interference, and security requirements. Due to the non-convexity of the problem, a Taylor expansion is used to simplify complex terms, and a Successive Lower bound Maximization (SLM) approach is proposed for efficient problem-solving. Simulation results demonstrate that the proposed scheme enhances communication efficiency while ensuring system security against eavesdropping.
In this paper \cite{chen2024jointJoint}, the authors explore combined beamforming with RIS in a secure ISAC system. The BS transmits interference signals alongside communication signals to enhance PLS and perform dual functions. Optimization focuses on the BS's active beamformer and the RIS's passive beamforming matrix to improve SR and radiation power. A local search method is used for the RIS phase shift matrix, while FP and SDP optimize the active beamforming matrix. Simulation results validate the effectiveness of the proposed methods in enhancing C\&S performance.



\begin{table*}[htbp]
\caption{Summary of Optimizing Weighted Secrecy Rate and Sensing Under Constraints Schemes}
\label{T:Weighted}
\centering
\begin{tabular}{|l|p{0.55cm}|p{1.3cm}|p{.5cm}|p{.5cm}|p{.5cm}|p{1.1cm}|p{1.8cm}|p{1cm}|p{1.8cm}|p{1.5cm}| p{2.5cm}| }
\hline
\multirow{3}{*}{\textbf{Ref.}}   & \multirow{3}{*}{\textbf{Year}} & \multicolumn{2}{c|}{\textbf{System Model}} & \multicolumn{4}{c|}{\textbf{Communication and Sensing Details}} & \multirow{3}{*}{\textbf{Eve's CSI}} & \multirow{3}{*}{\textbf{Methodology}}     & \multirow{3}{*}{\textbf{Opt. Variables}} & \multirow{3}{*}{\textbf{Objective}} \\ \cline{3-8}
& & \textbf{Scenario} & \textbf{\# (RISs)} & \textbf{\# (UEs)} & \textbf{\# (Eves)} & \textbf{Sensing Type} & \textbf{Sensing Metric} & & & & \\ \hline
\cite{jin2023ris} & 2023 &  RIS-DFRC  & S & S &  S &  Eve sensing &  Ambiguity function \& SR & Avail. & FSCSS-IM & Tx BFs and RIS PSs & To secure communication and improve radar sensing performance.\\ \hline

\cite{wei2024cramer} & 2024 & Semi-passive RIS-ISAC & S & S &  M &  Targets sensing & CRB & Avail. &  SCA, SDR, AO &  Tx BF, RIS PSs & To maximize SR and minimize sensing CRB. \\ \hline


\cite{song2024secure} & 2024 & RIS-ISAC & S & M &  S & Eve sensing & CRB, SNR  & Avail. & SLM, Taylor expansion & 
Tx BFs and RIS PSs & To maximize SNR and minimize CRB while ensuring security. \\ \hline
\cite{chen2024jointJoint} & 2024 & RIS-ISAC& S & M &  S &  Target
detection & max. achievable SR \& Radiation Power & Avail.  &  FP, SDP  & Active BF matrix and RIS PSs matrix & To enhance sensing accuracy and secure communication. \\ \hline


\end{tabular}
\end{table*}



\subsubsection*{Discussion} 
As summarized in Table \ref{T:Weighted}, this subsection examines WSR and sensing optimization in RIS-assisted ISAC, focusing on secure communication and target detection through beamforming, RIS phase tuning, and interference-based security measures. Studies such as \cite{jin2023ris, wei2024cramer, song2024secure, chen2024jointJoint} employ DFRC, semi-passive RIS, and symbol-level precoding to enhance security and sensing accuracy. \cite{jin2023ris} utilizes FSCSS-IM to integrate radar-aided beamforming for PLS and BER improvement. \cite{wei2024cramer} optimizes CRB minimization and SR maximization using SCA, SDR, and AO, assuming available CSI. \cite{song2024secure} applies SLM and Taylor expansion to improve SNR and CRB, while \cite{chen2024jointJoint} leverages FP and SDP for joint beamforming, ensuring secure transmission and effective sensing.

Despite advancements, limitations persist. Most studies assume ideal RIS settings and unlimited power, overlooking hardware constraints. Additionally, none consider imperfect CSI, which is unrealistic for real-world ISAC systems where accurate CSI is often unavailable. While balancing sensing and secrecy, they lack multi-user security models, limiting scalability. Furthermore, symbol-level precoding and interference-based security in adaptive beamforming remain underexplored for eavesdropping resistance. Future research should focus on hardware-efficient RIS deployment, scalable security models, and robust CSI estimation techniques to improve the practicality and resilience of ISAC systems.

\section{ARIS-based Advanced PLS Approaches in ISAC Systems} 
\label{sec:sectionIV} 
\begin{figure}
\centering
\includegraphics[width=0.40\textwidth]{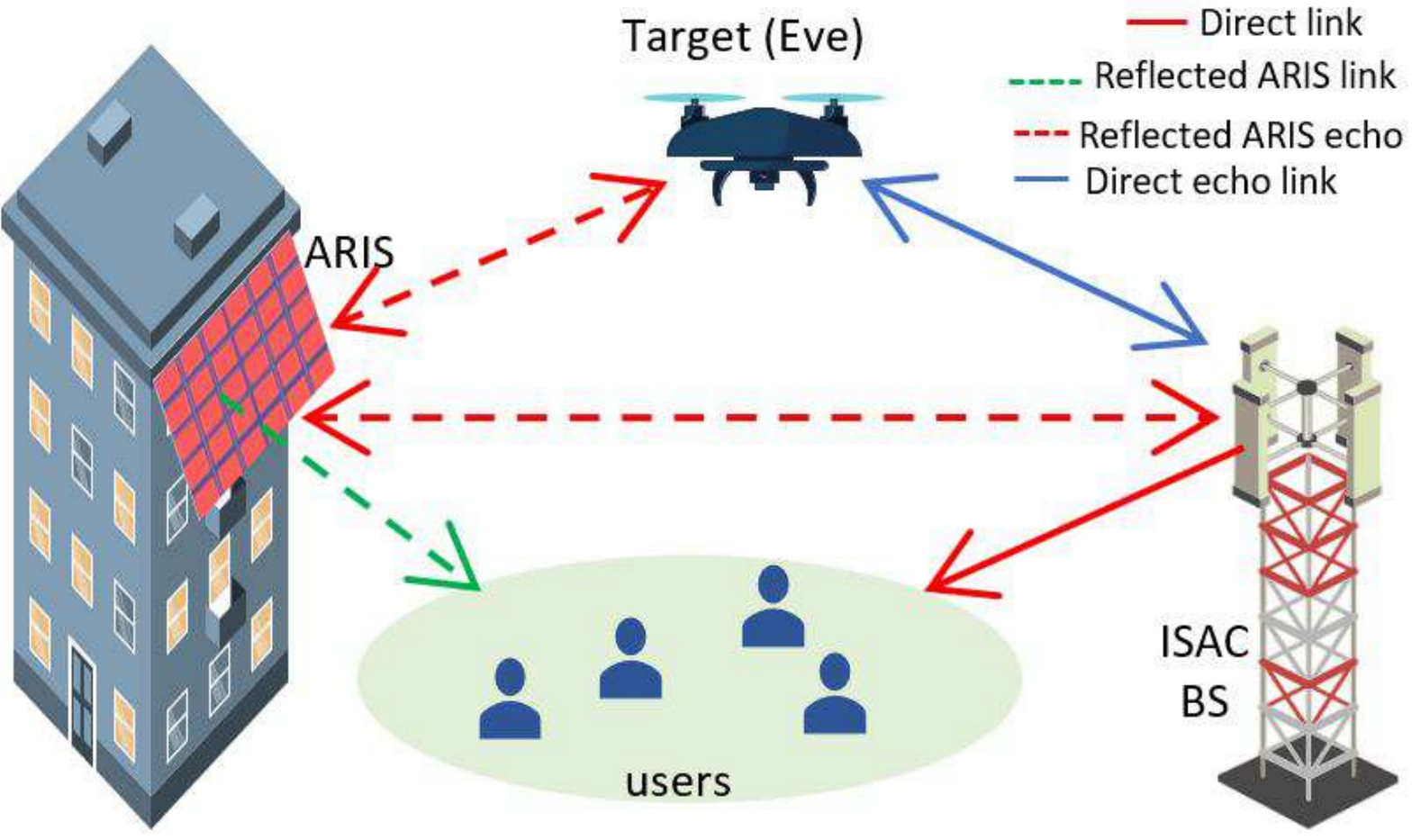}
\caption{Active-RIS enabled ISAC system with a malicious UAV.}
\label{fig:optimizing secracy rates}
\end{figure}

\begin{figure}
\centering
\includegraphics[width=0.40\textwidth]{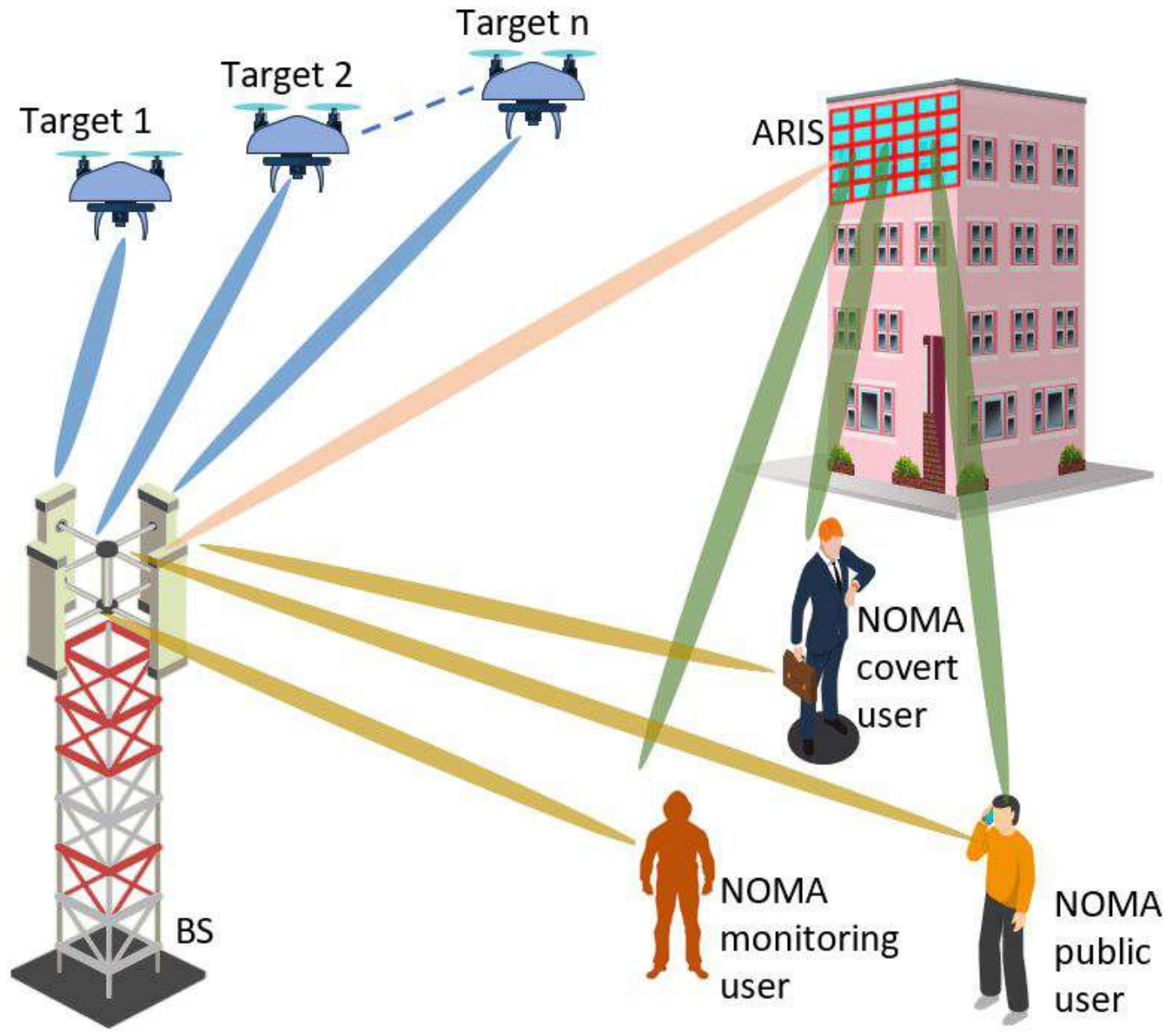}
\caption{ARIS-aided NOMA-ISAC system with a DFBS,  a warden  (monitoring) user, NOMA users (public and covert), and multiple moving targets for simultaneous communication and sensing.}
\label{fig:maximizing covert rate}
\end{figure}

This section discusses the advantages of employing ARIS-based PLS methods within ISAC systems for dynamic beamforming and amplification. These methods considerably bolster security, as illustrated in Fig. \ref{fig:optimizing secracy rates}, and enable covert communication, as demonstrated in Fig. \ref{fig:maximizing covert rate}, while also enhancing sensing capabilities.  ARIS offers superior control compared to passive RIS, resulting in increased resistance to eavesdropping through optimized beamforming and effective power management. The primary objective is to enhance both secrecy and communication rates while maximizing sensing SINR. This enhancement is achieved through methodologies such as AN, RSMA, NOMA-ISAC, and STAR-RIS designs. Advanced optimization techniques, including SDP, AO, and SCA, are employed to effectively balance the trade-offs between security and sensing. These advancements underscore the significant role of ARIS in improving security, sensing, and communication capabilities within future ISAC systems.
\vspace{-15pt}
\subsection{Optimizing Secrecy Rates Under Sensing and Other Constraints}
 In \cite{salem2022active}, the authors investigate the PLS of a Multi-User Multiple Input Single Output (MU-MISO) ISAC system operating under the threat of eavesdropping by a malicious UAV. To counter this, an ARIS is utilized to maximize the SR by jointly optimizing the radar receive beamformers, RIS RCs, and transmit beamformers at the ISAC BS, while ensuring a minimum radar detection SNR and meeting power constraints. Given the non-convexity of the problem, FP  and MM techniques are applied for efficient optimization. Simulation results confirm that ARIS markedly enhances security when compared to both passive RIS and RIS-free ISAC systems. While a passive RIS remains a viable option, achieving comparable performance necessitates a significantly larger surface area, particularly under low-power constraints.

The research presented here \cite{sun2024secure} investigates secure transmission in an ARIS-assisted Terahertz (THz) ISAC system, where delay alignment modulation is implemented at the BS. Considering the target as a potential Eve, the goal is to maximize the SR while ensuring minimum illumination power. Given the non-convex nature of this optimization problem, a novel algorithm is introduced. This algorithm iteratively optimizes both the BS transmit beamforming and the reflection coefficients of the ARIS using MM and SDR techniques. Simulations demonstrate the effectiveness of the proposed approach in enhancing secure transmission. Prior studies, such as \cite{salem2022active, sun2024secure}, have primarily focused on either Eve detection or target sensing. However, the following works consider both aspects simultaneously. For example, \cite{zhang2024secure-arxiv} examines a DFRC system with an ARIS and a potential Eve, aiming to maximize SR by jointly optimizing the beamforming matrix at the DFRC-BS and the reflection coefficients of the ARIS, while ensuring the SINR constraint for radar echo and adhering to power consumption limits at both the DFRC-BS and ARIS.
To solve this SR-maximization problem, an AO approach utilizing SDR and MM is applied. First, SDR and SCA transform the subproblems into more tractable forms, followed by MM to derive a concave surrogate function for iterative optimization. Simulation results demonstrate that ARIS effectively mitigates the impact of multiplicative fading and surpasses passive RIS in both secure data transmission and radar sensing performance. Unlike the works in \cite{salem2022active, sun2024secure, zhang2024secure-arxiv}, the authors in
\cite{salem2024robust} consider imperfect CSI 
 and integrates RSMA with ISAC poses security issues related to potential Eves. The distinction between public and private channels heightens eavesdropping threats due to vulnerabilities in the public channel. This investigation presents a strategy utilizing ARIS beamforming and AN to improve the security of RSMA-based ISAC. An EPSR is determined through a mathematical estimation of Eve's average channel gain. An optimization problem aims to maximize the minimum EPSR while fulfilling the conditions for ergodic common SR, radar sensing, and RIS power limits. A new optimization approach is proposed to solve this non-convex challenge by iteratively optimizing the transmit beamforming matrix for both common and private streams, integrating rate splitting, AN, the RIS RC, and the radar receive beamformer. The results from simulations indicate that the proposed method significantly outperforms existing benchmarks, demonstrating its potential for improving the security landscape in RSMA-based ISAC systems. 

\vspace{-15pt}
\subsection{Maximizing Covert Rates Under Sensing and Other Constraints} Unlike the works in Subsection \ref{sec:sectionIV}A, which focus on improving SRs under various constraints, these studies emphasize covert rate maximization while ensuring sensing and other constraints are met.
In \cite{zhu2024active}, ISAC utilizing NOMA allows radar sensing and communication in the same spectrum, which raises privacy and security concerns due to uncontrolled wireless transmission. To enhance secure communication in this NOMA-based ISAC setup, an ARIS is introduced to improve SRs. Covert rate optimization is achieved through joint beamforming at the BS and reflection beamforming at the RIS, considering two scenarios: with a Dedicated Sensing Signal (w-DSS) and without (w/o-DSS). An AO algorithm is proposed to address the challenges of maximizing the nonconcave secret rate. Numerical results demonstrate that the ARIS-assisted NOMA-ISAC system significantly surpasses both passive-RIS and non-RIS configurations in SR, effectively integrating secure communication with sensing capabilities. Similarly, in \cite{liuexploiting}, the investigation delves into ARIS for secure communication within a millimeter-wave ISAC network characterized by limited block length. The configuration involves a BS (Alice) that conducts single-target sensing while simultaneously transmitting private messages to a user (Bob). ARIS improves target reflection and interferes with Bob's signal by altering radio propagation and decreasing multiplicative fading. The minimum Detection Error Probability (DEP) for the target is formulated in closed form through a large system analytical method. A combined beamforming technique is introduced to bolster Bob's secure communication, optimizing transmission from Alice and RIS's reflection beamforming to augment the secret rate. An algorithm incorporating AO and SCA is developed to address these complexities. Similarly, \cite{liu2025active} addresses various security threats, specifically focusing on eavesdropping and detection attacks, by incorporating an active STAR-RIS for unified protection. A multi-antenna base station simultaneously performs target sensing while communicating with Secrecy Users (SUs) and Covert Users (CUs), the target inadvertently acts as an eavesdropper, with an entity named Willie attempting to detect the transmissions. To counter these threats effectively, a joint strategy for secrecy and covert communication is proposed. The BS utilizes channel fading discrepancies to reduce data leakage from SUs while employing Gaussian signaling techniques to obscure transmissions aimed at CUs.

Using Pinsker’s inequality and large-system analysis, bounds on Willie’s DEP are established. To balance radar sensing and secure communication, the BS beamforming, filtering, and STAR-RIS transmission/reflection are jointly optimized to maximize secrecy and covert rates while satisfying SNR, DEP, and power constraints. An iterative algorithm leveraging SCA, SDR, and rank-one relaxation is developed for this optimization.
Simulations confirm active STAR-RIS effectively mitigates multiplicative fading and significantly outperforms baseline schemes in defending against hybrid attacks.

\subsection{Maximizing  Sensing SINR Under Communication and Secrecy Constraints}

This paper \cite{peng2024robust} investigates security in an active RIS-aided RSMA ISAC system, where targets act as potential Eves under imperfect CSI. A robust active beamforming strategy is proposed to maximize radar SINR while meeting SOP constraints. Due to non-convexity, a suboptimal iterative algorithm using SDR and Taylor expansion is introduced to alternately optimize two subproblems. Simulations confirm the AO algorithm enhances secure C\&S, outperforming conventional approaches. In a different research work \cite{kumar2025beamforming}, the authors optimize beampattern gain in a secure passive RIS/active RIS-aided ISAC system, considering SINR and information leakage constraints. A SCA-based approach is proposed to jointly optimize beamforming, updating all variables simultaneously unlike AO-based methods.
Simulations confirm the proposed method outperforms penalty-based AO for passive RIS-aided ISAC, while ARIS achieves higher beampattern gain under the same power budget. Complexity analysis and convergence proof validate its efficiency.

\subsubsection*{Discussion} As summarized in Table \ref{T:Active},  ARIS significantly enhances ISAC security by not only reflecting but also amplifying signals, mitigating multiplicative fading, and improving SR, sensing SINR, and covert communication performance \cite{salem2022active, sun2024secure, zhang2024secure-arxiv, salem2024robust, zhu2024active, liuexploiting, liu2025active, peng2024robust, kumar2025beamforming}. Rather than relying on Passive RIS, which necessitates larger surfaces to achieve comparable performance under low-power constraints, ARIS optimizes secrecy and sensing trade-offs more effectively \cite{salem2022active, sun2024secure}. 

However, ARIS introduces new challenges, including higher power consumption, hardware complexity, and heat dissipation, which most works fail to fully address \cite{salem2024robust, peng2024robust}. Additionally, optimization methods in ARIS-based ISAC, such as AO, SDR, MM, and SCA, are often offline and lack real-time adaptability in dynamic environments \cite{sun2024secure, zhang2024secure-arxiv, salem2024robust}. While ARIS enhances covert communication by actively manipulating radio propagation \cite{zhu2024active, liuexploiting, liu2025active}, most studies lack multi-user security models and fail to consider adversarial learning-based attacks. For sensing-oriented security, ARIS-based methods achieve higher SINR and improved target detection \cite{peng2024robust, kumar2025beamforming}, but robust interference-aware optimization remains underexplored. Furthermore, several studies assume perfect or available CSI \cite{salem2022active, sun2024secure, zhang2024secure-arxiv, liuexploiting, kumar2025beamforming}, which is unrealistic in practical eavesdropping scenarios. Future research directions should focus on developing power-efficient ARIS designs to address energy constraints, integrating ML-based optimization for real-time security adaptation, introducing multi-user covert security frameworks for large-scale ISAC networks, and leveraging probabilistic CSI estimation with adversarial learning-based techniques to enhance security in practical deployments.

\vspace{-10pt}
\section{Lessons learned, open issues and future research directions }\label{sec:sectionV} This subsection explores key insights gained, unresolved challenges, and future research directions in RIS-based PLS for ISAC.
\begin{table*}[htbp]
\caption{Summary of ARIS-Based Advanced PLS Approaches in ISAC Systems}
\label{T:Active}
\centering
\begin{tabular}{|p{0.55cm}|l|p{0.55cm}|p{1.3cm}|p{.5cm}|p{.5cm}|p{.5cm}|p{1.1cm}|p{1.2cm}|p{1cm}|p{1.8cm}|p{1.5cm}| p{2.5cm}| }
\hline
\multirow{3}{*}{\textbf{{Cat.}}} & \multirow{3}{*}{\textbf{Ref.}} & \multirow{3}{*}{\textbf{Year}} & \multicolumn{2}{c|}{\textbf{System Model}} & \multicolumn{4}{c|}{\textbf{Communication and Sensing Details}} & \multirow{3}{*}{\textbf{Eve's CSI}} & \multirow{3}{*}{\textbf{Methodology}} & \multirow{3}{*}{\textbf{Opt. Variables}} & \multirow{3}{*}{\textbf{Objective}} \\ \cline{4-9}
& & & \textbf{Scenario} & \textbf{\# (RISs)} & \textbf{\# (UEs)} & \textbf{\# (Eves)} & \textbf{Sensing Type} & \textbf{Sensing Metric} & & & & \\ \hline

& \cite{salem2022active} & 2022 & MU-MISO-ISAC & S & M & S & Eve. detection  & min. radar detection SNR & Avail. & FP, MM & Tx BF, radar Rx BF, ARIS RCs & To maximize achievable SR. \\ \cline{2-13}
\multirow{3}{*}{\rotatebox{90}{\shortstack{Maximizing Secrecy Rate  \\ Schemes}}} 
& \cite{sun2024secure} & 2024 & ARIS-THz-ISAC & S & S & S & Target sensing & min. illumination power & Avail. & AO, MM, SDR & Tx BF, ARIS RCs & To maximize SR, ensure target illumination and sensing accuracy. \\ \cline{2-13}

& \cite{zhang2024secure-arxiv} & 2024 & ARIS--DFRC & S & S & S & Eve. sensing \& target detect & Radar SINR & Perfect & AO, SDR, SCA, MM & Tx BF, ARIS RCs & To maximize SR, ensure radar SINR and power constraints. \\ \cline{2-13}

& \cite{salem2024robust} & 2024 & ARIS-RSMA-ISAC & S & M & S & Target  \& Eve. sensing & max.min EPSR \& ECSR, min.sensing SNR & Imperfect & SCA, MM & Tx BF, Rate Splitting Coefficients, AN, RIS RC, radar Rx BF & To maximize EPSR while ensuring secrecy, sensing accuracy and QoS. \\ \hline

& \cite{zhu2024active} & 2024 & ARIS-NOMA-ISAC & S & M & S & Targets sensing  & CRB & Statistical & Penalized dinkelbach transformation and AO & BS Tx BF \& ARIS RCs & To maximize covert rate while ensuring sensing accuracy and QoS. \\ \cline{2-13}
\multirow{3}{*}{\rotatebox{90}
{\shortstack{Maximizing Covert Rate \\ Schemes}}} 
& \cite{liuexploiting} & 2024 & ARIS--mmWave-ISAC & S & S & S & Target sensing & min. DEP, sensing SINR & Avail. & AO, SCA, Large System Analytical Method & Tx BF, reflection BF & To maximize covert rate, ensure minimum DEP and sensing SINR. \\ \cline{2-13}

& \cite{liu2025active} & 2025 & Active STAR-RIS ISAC & S & M & M & Eve   detection & Echo SNR, DEP & Avail. & SCA \& SDR& Tx BF, Receive Filter, RIS Tx \& Reflect BF & To maximize covert and sum rates while ensuring radar sensing constraints and mitigating hybrid attacks. \\ \hline \color{black}

\multirow{2}{*}{\rotatebox{90}{\shortstack{Maximizing Sensing \\ SINR  Schemes}}}
& \cite{peng2024robust} & 2024 & ARIS-RSMA-ISAC & S & M & M & Targets sensing  & Sensing SINR & Imperfect & SDR, Taylor Expansion, AO & Tx BF, ARIS RCs & To maximize radar SINR, ensure secure communication and SOP constraints. \\ \cline{2-13}

& \cite{kumar2025beamforming} & 2025 & ARIS-ISAC & S & M & S & Target sensing & SINR \& Beampattern gain & Perfect & SCA, SDR & BF vector, PS Vector, Power alloc. & To maximize beampattern gain at the eavesdropping target while ensuring communication SINR and limiting interference. \\ \hline


\end{tabular}
\end{table*}

\vspace{-15pt}
\subsection{Lessons Learned}
\begin{itemize}

 \item  
 Passive {RIS} improves security by modifying the wireless environment, but its effectiveness is limited by phase shift constraints. In contrast, {ARIS} actively optimizes beamforming, offering better control over secrecy and covert communication, but at the cost of increased energy consumption and system complexity.

 \item  There exists a trade-off between security, communication efficiency, and sensing performance. Enhancing SR or covert communication often leads to reduced sensing accuracy or system throughput, necessitating joint optimization strategies that balance these conflicting objectives. Additionally, {PLS} techniques such as secure beamforming, AN, and jamming strengthen security but are insufficient against sophisticated attacks. Combining {RIS}-{PLS} with cryptographic techniques, AI-based anomaly detection, and network-layer security mechanisms can enhance resilience.

 \item Despite its theoretical potential, hardware limitations impact the practical deployment of {RIS}-assisted {PLS}. Factors such as signal attenuation, phase noise, and limited phase resolution can degrade system performance, highlighting the need for energy-efficient {RIS} designs and scalable implementations. 
 
  \item Furthermore, channel estimation remains a critical challenge, as accurate {CSI} is essential for optimizing security and sensing in {ISAC} systems. However, conventional estimation methods often fall short due to the passive nature and large array size of {RIS}, making AI-driven and hybrid approaches necessary for improved accuracy.

 \item Intelligent and adaptive {RIS}-{PLS} solutions will be crucial for securing {ISAC} in {6G} networks. Future research should focus on autonomous {RIS} configurations that dynamically adjust phase shifts and beamforming strategies in response to real-time security threats. The integration of ML and advanced optimization algorithms will further enhance system resilience, making {RIS}-{PLS} a scalable and efficient solution for next-generation {ISAC} networks.
\end{itemize}
\vspace{-12pt}
\subsection{Open Issues and Future Research Directions}
\begin{itemize}
     \item [$\bullet$] \textbf{Synergy with AI:} The integration of ISAC with AI possesses significant promise for augmenting security through the facilitation of real-time threat identification and response. AI-driven algorithms can evaluate sensory and communication data to find anomalies, detect cyber threats, and anticipate potential security breaches prior to their occurrence. AI use in ISAC security can fortify multiple sectors, such as smart cities, supply chain security, agriculture, intelligent transportation, environmental monitoring, and public safety, by proactively mitigating security weaknesses and augmenting resilience against cyber threats.Future research should concentrate on creating AI-driven systems that can analyze sensing and communication data to find anomalies, detect cyber threats, and proactively predict potential security breaches.

     \item [$\bullet$] \textbf{RIS Power Autonomy:} Ensuring power autonomy in RIS remains a significant challenge, as traditional RIS designs rely on external power sources such as electrical grids or batteries. While RIS consumes less energy than active relays, its power requirements increase with the number of elements, making energy-efficient operation critical for large-scale deployments in ISAC systems. To address this, energy harvesting solutions such as RF energy harvesting and solar-assisted RIS have been explored. RF energy harvesting allows RIS to capture ambient radio waves and convert them into power, but its efficiency is limited by energy availability and conversion rates, making it insufficient for high-power applications. Solar-assisted RIS offers a more sustainable alternative, enabling both signal processing and power generation without additional infrastructure. However, environmental factors such as sunlight availability and efficiency constraints pose further challenges. Achieving energy autonomy in RIS-based PLS for ISAC is vital for secure operation in energy-constrained environments. Future research should focus on hybrid energy-harvesting architectures that optimize power management while ensuring security and sensing efficiency.

   \item [$\bullet$] \textbf{Quantum-Secure RIS-Assisted ISAC:}
With the progression of quantum computing, conventional encryption techniques employed in ISAC systems encounter considerable security vulnerabilities. Consequently, forthcoming research should concentrate on amalgamating quantum-resistant security protocols with RIS-assisted ISAC to guarantee enduring defense against quantum assaults.
Furthermore, subsequent investigations should focus on the incorporation of hybrid classical-quantum security frameworks within RIS-assisted ISAC. As the shift to entirely quantum-secure networks may need time, it is essential to design hybrid security frameworks that integrate conventional cryptographic methods with novel quantum-safe strategies. This involves employing RIS to enhance secure key distribution, authentication, and encryption in both classical and quantum-secure ISAC contexts.


\item [$\bullet$] \textbf{Channel Estimation Challenges :} As the number of RIS elements escalates and near-field propagation effects intensify, achieving real-time and precise channel estimate in RIS-assisted ISAC systems becomes progressively challenging. The complexity stems from the multitude of channel coefficients and the swift dynamic fluctuations in high-frequency settings. This presents a considerable difficulty for secure ISAC operations, since erroneous channel estimation can result in diminished sensing accuracy and heightened susceptibility to eavesdropping or jamming assaults. Furthermore, the overhead linked to pilot transmission and training becomes burdensome, rendering it unfeasible for terminals to efficiently monitor all potential cascaded channels. Utilizing a structured codebook technique, rather than exhaustive search-based near-field beam training, can markedly decrease pilot transmission overhead while facilitating more efficient and precise CSI estimation. Future research should concentrate on refining RIS-based channel estimate methodologies to improve security and performance in ISAC networks.

\item [$\bullet$] \textbf{False RIS Attack:}
Security challenges in RIS-assisted PLS for ISAC require advanced PLS mechanisms to counter emerging threats, particularly false RIS attacks, where adversaries impersonate legitimate RIS to compromise the network. Future research should focus on secure authentication methods, including wired (PoE-based), wireless (SIM-based), and meta-cryptography-based authentication to ensure robust identity verification.
Additionally, secure RIS control strategies need exploration, optimizing BS-controlled and UE-controlled operations. Integrating secure feedback mechanisms will further enhance authentication and prevent unauthorized access \cite{10833623}.
To strengthen RIS-PLS in ISAC, future studies should develop AI-driven anomaly detection, hybrid cryptographic solutions, and adaptive PLS frameworks to ensure secure, efficient, and resilient RIS-assisted ISAC in 6G networks.

\end{itemize}
\vspace{-10pt}
\section{Conclusion} \label{sec:sectionVI}This survey examined the integration of RIS and PLS techniques within ISAC systems, emphasizing their significance for secure communication in future 6G networks. Initially, it summarized the fundamental concepts of RIS, PLS, and ISAC, differentiating between D-RIS and BD-RIS architectures and highlighting their operational modes along with practical implications. A detailed analysis was presented for passive RIS and ARIS paradigms, illustrating their unique roles in balancing security, covert communication, and sensing performance.
Key observations identified inherent trade-offs among security enhancement, communication throughput, and sensing accuracy. While passive RIS technologies offered effective environmental control for security advancements, their performance was limited by restricted phase-shifting capabilities. In contrast, ARIS approaches facilitated advanced beamforming to optimize secrecy and sensing performance; however, they introduced increased complexity and higher energy requirements. Challenges in practical implementation, such as hardware imperfections and accurate CSI acquisition, were underscored as crucial issues. Addressing these challenges is essential for the successful realization of secure, robust, and scalable RIS-assisted ISAC in 6G and beyond. Progress in these areas is vital for achieving secure, robust, and scalable RIS-assisted ISAC in 6G and beyond. 
\vspace{-5pt}


\ifCLASSOPTIONcaptionsoff
  \newpage
\fi

%

\bibliographystyle{IEEEtran.bst}
\bibliography{main}

%






\end{document}